\documentclass[a4paper,fleqn]{cas-dc}

\usepackage[authoryear]{natbib}

\graphicspath{{./images/}}
\DeclareGraphicsExtensions{.pdf,.jpeg,.png}
  
\usepackage{acro}
\DeclareAcronym{sme}{
  short = SME ,
  long  = small and medium-sized enterprise ,
  class = abbrev
}
\DeclareAcronym{ism}{
  short = ISM ,
  long  = Information Security Management ,
  class = abbrev
}
\DeclareAcronym{isrm}{
  short = ISRM ,
  long  = Information Security Risk Management ,
  class = abbrev
}
\DeclareAcronym{isms}{
  short = ISMS ,
  long  = Information Security Management System ,
  class = abbrev
}
\DeclareAcronym{pii}{
  short = PII ,
  long  = Personal Identifiable Information ,
  class = abbrev
}
\DeclareAcronym{tam2}{
  short = TAM2 ,
  long  = Extended Technology Acceptance Model ,
  class = abbrev
}
\DeclareAcronym{tam}{
  short = TAM ,
  long  = Technology Acceptance Model ,
  class = abbrev
}
\DeclareAcronym{ucit}{
  short = UCIT ,
  long  = Universal Constructive Instructional Theory ,
  class = abbrev
}
\DeclareAcronym{elise}{
  short = ELISE ,
  long  = Efficient Learning- and Information-System-Evaluation ,
  class = abbrev
}
\DeclareAcronym{iqr}{
  short = IQR ,
  long  = Interquartile Range ,
  class = abbrev
}
\DeclareAcronym{ciso}{
  short = CISO ,
  long  = Chief Information Security Officer,
  class = abbrev
}
\DeclareAcronym{cio}{
  short = CIO ,
  long  = Chief Information Officer ,
  class = abbrev
}
\DeclareAcronym{cto}{
  short = CTO ,
  long  = Chief Technology Officer ,
  class = abbrev
}
\DeclareAcronym{cissp}{
  short = CISSP ,
  long  = Certified Information Systems Security Professional ,
  class = abbrev
}
\DeclareAcronym{cism}{
  short = CISM ,
  long  = Certified Information Security Manager ,
  class = abbrev
}
\DeclareAcronym{itil}{
  short = ITIL ,
  long  = Information Technology Infrastructure Library ,
  class = abbrev
}
\DeclareAcronym{fmea}{
  short = FMEA ,
  long  = Failure Mode and Effect Analysis ,
  class = abbrev
}
\DeclareAcronym{cca}{
  short = CCA ,
  long  = Cause and Consequence Analysis ,
  class = abbrev
}
\DeclareAcronym{fta}{
  short = FTA ,
  long  = Fault Tree Analysis ,
  class = abbrev
}
\DeclareAcronym{isram}{
  short = ISRAM ,
  long  = Information Security Risk Analysis Method ,
  class = abbrev
}
\DeclareAcronym{swift}{
  short = SWIFT ,
  long  = Structured What If Technique ,
  class = abbrev
}
\DeclareAcronym{pha}{
  short = PHA ,
  long  = Preliminary Hazard Analysis ,
  class = abbrev
}
\DeclareAcronym{hazop}{
  short = HAZOP ,
  long  = Hazard and Operability Study ,
  class = abbrev
}
\DeclareAcronym{ram}{
  short = RAM ,
  long  = {Reliability Availability, Maintainability Analysis} ,
  class = abbrev
}
\DeclareAcronym{tpb}{
  short = TPB ,
  long  = Theory of Planned Behavior ,
  class = abbrev
}
\DeclareAcronym{pmt}{
  short = PMT ,
  long  = Protction Motivation Theory ,
  class = abbrev
}
\DeclareAcronym{grc}{
  short = GRC ,
  long  = {Governance, Risk Management and Compliance} ,
  class = abbrev
}
\DeclareAcronym{rmf}{
  short = RMF ,
  long  = {Risk Management Framework} ,
  class = abbrev
}

\usepackage{multicol}
\usepackage{multirow}
\usepackage{xspace}
\usepackage{tabularx}

\newcommand{\shortquote}[1]{\textit{``#1''}}

\usepackage{balance}

\usepackage{tikz}

\newcommand*\circledb[1]{\tikz[baseline=(char.base)]{
            \node[shape=circle,fill,inner sep=2pt] (char) {\textcolor{white}{#1}};}\hspace{1pt}}

\begin{document}
\let\WriteBookmarks\relax
\def\floatpagepagefraction{1}
\def\textpagefraction{.001}
\shorttitle{Risk Management Practices in Information Security: Exploring the Status Quo in the DACH Region}
\shortauthors{Michael Brunner et~al.}

\title [mode=title]{Risk Management Practices in Information Security}                      
\title [mode=sub]{Exploring the Status Quo in the DACH Region}                      

\author{Michael Brunner}
\cormark[1]
\ead{michael.brunner@uibk.ac.at}
\credit{Conceptualization, Methodology, Investigation, Data curation, Formal analysis, Writing - Original draft preparation, Visualization, Project administration}

\address{University of Innsbruck, Department of Computer Science, Technikerstrasse 21A, 6020 Innsbruck, Austria}

\author{Clemens Sauerwein}
\ead{clemens.sauerwein@uibk.ac.at}
\credit{Conceptualization, Methodology, Investigation, Validation, Writing - Original draft preparation}

\author{Michael Felderer}
\ead{michael.felderer@uibk.ac.at}
\credit{Conceptualization, Resources, Writing - Review \& Editing, Supervision}

\author{Ruth Breu}
\ead{ruth.breu@uibk.ac.at}
\credit{Writing - Review \& Editing, Supervision, Funding acquisition}
 
\cortext[cor1]{Corresponding author}

\begin{abstract}
Information security management aims at ensuring proper protection of information values and information processing systems (i.e. assets). Information security risk management techniques are incorporated to deal with threats and vulnerabilities that impose risks to information security properties of these assets. 
This paper investigates the current state of risk management practices being used in information security management in the DACH region (Germany, Austria, Switzerland).  
We used an anonymous online survey targeting strategic and operative information security and risk managers and collected data from 26 organizations. 
We analyzed general practices, documentation artifacts, patterns of stakeholder collaboration as well as tool types and data sources used by enterprises to conduct information security management activities.
Our findings show that the state of practice of information security risk management is in need of improvement. Current industrial practice heavily relies on manual data collection and complex potentially subjective decision processes with multiple stakeholders involved. Dedicated risk management tools and methods are used selectively and neglected in favor of general-purpose documentation tools and direct communication between stakeholders.
In light of our results we propose guidelines for the development of risk management practices that are better aligned with the current operational situation in information security management.
\end{abstract}




\maketitle


\section{Introduction}\label{sec:introduction}

Information security is concerned with the protection of information regarding confidentiality, integrity and availability. With the advent of tighter regulatory demands regarding information security (such as the EU GDPR and NIS Directive) and increasing customer demands, enterprises are forced to establish measures to ensure the information security of their valuable assets. This especially applies to enterprises working with protected \ac{pii}, payment data or other sensitive information. Establishing and operating an \ac{isms} has become the tool of choice to systematically deal with information security risks. \acp{isms} provide actionable sets of requirements, policies, guidelines and process definitions to aid organizations in their quest to comply with their individual information security goals. \ac{isrm} is a vital part of any \ac{isms} ensuring that information security risks are systematically identified, analyzed and treated in accordance with an organization’s risk appetite.

Over the last years the overall greater need to systematically comply with information security goals has supported the raise of a plethora of tools and methods to support all ranges of information security and information security risk management activities. For example, there has been substantial research in the state of practice of specific \ac{ism} and \ac{isrm} practices. The application of information security policies \citep{Bulgurcu:2010ft,Fulford:2003cv,Sommestad:2014fs} or the automation of security controls to mitigate identified information security risks \citep{Montesino:2011kb,Aguirre:2012dz,Tracy:2007jq} as well as the support of specific risk identification techniques \citep{Beckers:2015ju,Sommestad:2013gu,Panda:2009vo} have been at the center of attention for more than a decade. Still, the general practical application and dissemination of these approaches as well as their integration in existing organizationally established \acp{isms} or \ac{isrm} settings, is hard to gauge \citep{Wangen:2013wb}. Part of this situation is due to enterprises not readily disclosing their \ac{isrm} practices especially when past missteps might have disrupted their \ac{ism} activities. Furthermore, many of the approaches proposed from academia target large-sized companies and specific information security pains they already face. Thus, they might impose higher-than-acceptable costs for other enterprises. The resulting inability of enterprises to reliably estimate the cost-benefit ratio of these highly specialized approaches might prevent them to justify their implementation, especially in light of their own pressing information security needs, their current state of practice and their budgetary boundaries. Consequently, researchers would benefit from a better understanding of the current situation regarding \ac{ism} and \ac{isrm} to better tailor their approaches to a broader audience and ensure a more general applicability of their results.

While \ac{ism} is generally considered a standardized discipline with explicit \ac{isms} standards such as ISO 27001 \citep{ISO:2013wu} or the BSI IT Baseline Protection Methodology \citep{bsi:2017ms}, the actual application within enterprises may vary greatly thanks to tolerances these standards allow. Apart from roles, workflow descriptions (which are in case of ISO 27001 rather abstract) and general requirements (which predominantly state the desired result, not the means to achieve it) there are no generally agreed upon tools or methods for conducting specific risk management activities within \acp{isms}. While standards do provide best-practices it is up for organizations to choose the most fitting ones for themselves, which in turn leads to highly heterogeneous \acp{isms} even if they are certified by the same standard. 

Research regarding the state of risk management practices applied within \acp{isms} is rather limited with prior publications either narrowly addressing specific aspects, focusing exclusively on the examinations of management practices, or investigating singular use cases (cf. Section~\ref{sec:background}). \cite{Wangen:2013wb} further illustrate the overall lack of good empirical research in the area of \ac{isms} and \ac{isrm}. Detailed studies covering \ac{isrm} practices of multiple enterprises including workflows, stakeholder collaboration and tool usage are not available. The goal of this research is therefore to evaluate the current practice and to identify potential shortcomings in \ac{ism} workflows, especially regarding the management of information security risks. The study at hand lays the foundation to address the following research objectives: 
\begin{itemize}
	\item Gain comprehensive understanding of the current state of practice of risk analysis used in \acl{ism}
	\item Improve the current organizational information security risk analysis practices
	\item Identify potential means for automatization in current risk analysis approaches applicable within \ac{isms} settings.
\end{itemize}
The scope of this exploratory investigation are enterprises that have either implemented or plan to implement an \ac{isms} to ensure that all participants apply \ac{isrm} on a broader scale as part of enterprise-wide information security management practices. Thus, we are not interested in smaller-scope, independent, non-information-security-centric risk management activities. Our study focuses on enterprises operating in the DACH region (Germany, Austria, Switzerland) primarily due to the EU GDPR~\citep{eu:2016gdpr} and NIS Directive~\citep{eu:2016nis} taking effect during our investigation period. We therefore presumed an increased organizational interest in information security and privacy in that geopolitical area.

With this paper we contribute (1) the design of a detailed survey to evaluate the current state of risk management practices conducted as part of organizations \acp{isms} together with (2) the study results and analysis for the DACH region as well as (3) the deduction of potential points for improvement in \ac{isrm} practices. Our findings will be further used to enhance the tool-supported and continuous \ac{ism} framework ADAMANT \citep{Brunner:2019fb,Brunner:2018wx,Brunner:2017by}.

The remainder of this paper is structured as follows. Sections~\ref{sec:background} and \ref{sec:relatedwork} describe the background and related work of our research. Section~\ref{sec:method} presents the applied research method and the developed survey instrument. The results of our exploratory survey are presented in Section~\ref{sec:results} before we discuss our conclusions and recommendations in Section~\ref{sec:conclusions}. We conclude this paper with a summary and outlook on future work in Section~\ref{sec:summary}.

\section{Background}\label{sec:background}

In this section we will present background to our study, mainly \ac{isms} and \ac{isrm} together with relevant standards, frameworks, and research resources. We will further present the conceptual model developed to guide our research and survey design.

\subsection{\acl{ism} and \aclp{isms}}\label{sec:background:ism}

While competing definitions for information security can be found, a commonly accepted one is provided by \cite{Whitman:2011kx}: \shortquote{Information security is the protection of information and its critical elements, including the systems and hardware that use, store, and transmit that information}. \ac{ism} consequently deals with the implementation and monitoring of an organization's desired information security level. An \ac{isms} is the management tool composed of interrelated and interacting organizational elements (policies, processes, roles, etc.) that supports the preservation of the confidentiality, integrity and availability of information values and information systems~\citep{ISO:2013wu}. These information values and information processing systems are commonly referred to as assets and managing an inventory of all relevant assets is a fundamental requirement for any given \ac{isms}. \ac{isrm} techniques are applied to systematically identify security risks of these assets, to analyze and evaluate them and to find proper means to treat the corresponding risks to information security. 

\begin{figure*}
\begin{center}
\includegraphics[width=.79\textwidth]{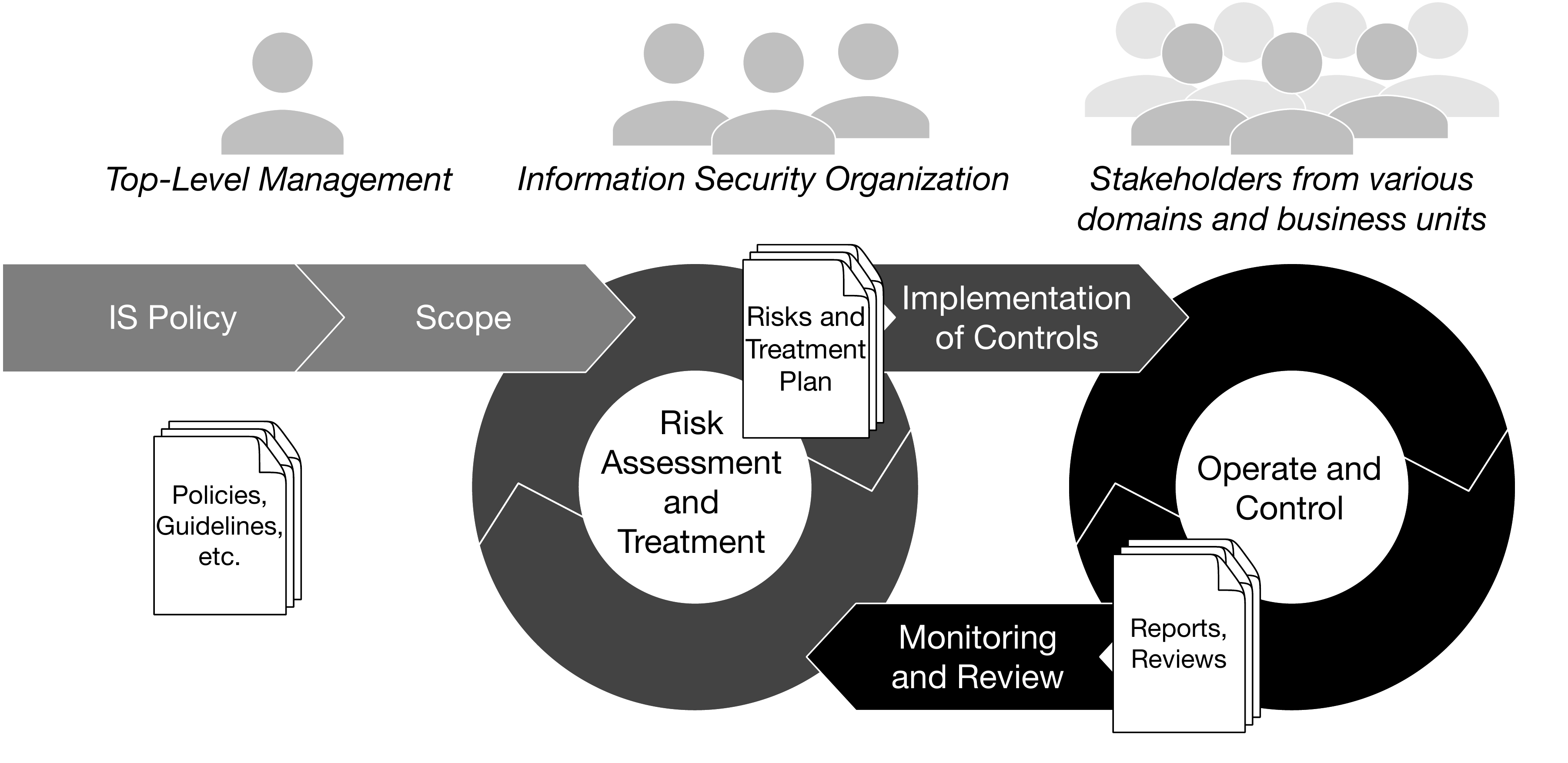}
\caption{Generalized \ac{isms} process}
\label{fig:isms-process}
\end{center}
\end{figure*}

In Figure~\ref{fig:isms-process} we present a unified, general mode of operation at the heart of any \ac{isms} derived from relevant standards and best practices. Top management will typically set the overarching goals by defining a strategic information security policy and the scope of the \ac{isms}. A broader set of stakeholders will then be responsible for operationalizing the outset goals by conducting a risk analysis, selecting appropriate counter measures to reduce the risk to the agreed level and subsequently implementing and operating them. This all is conducted as part of a continuous improvement cycle with reporting to top-level management and readjustment when necessary.

Recognized \ac{ism} standards and best practices are the ISO 27k family of standards~\citep{ISO:2013wu,ISO:2011wd}, the BSI IT Baseline Protection Methodology~\citep{bsi:2017ms}, the NIST \ac{rmf}~\citep{NIST:sp-800-37-r2} and COBIT \citep{cobit5}. While conceptually slightly different, these standards follow the described mode of operation and provide enterprises with useful guidelines. The actual implementation of said processes or the methods used for certain \ac{ism} and \ac{isrm} activities, however, are not mandated. The ISO 27001 standard, for example, list requirements that organizations risk assessment processes must meet (cf. ISO 27001, Section 6.1.2), but leaves the actual decision of the risk assessment methodology or the design of the process itself open. The BSI Baseline Protection Methodology as well as the NIST \ac{rmf} on the other hand mandate a more concrete risk assessment strategy starting with (1) a structure analysis to create the asset documentation, (2) the determination of protection requirements for each asset or group of assets (classification), (3) the implementation of according baseline security measures with the potential of (4) conducting dedicated a risk analysis for sensitive areas. However, the actual methodology used for the latter dedicated risk analysis is not mandated.

Academic \ac{ism} approaches and frameworks tend to partially integrate research results in common standards, often covering only selected parts or individual management activities. ISMS-CORAS~\citep{Beckers:2014gd} and subsequent work by \cite{Beckers:2015ju} explores the integration the model-driven CORAS~\citep{Lund:2011ip} risk management method into ISO 27001 compliant \acp{isms} for a smart grid scenario. Another model-driven approach incorporating enterprise architecture models is the Cyber Security Modeling Language (CySeMoL) \citep{Sommestad:2013gu}. Research in automation of \ac{ism} mainly considers the automation of security controls by applying heterogeneous sets of technical solutions to different domains of security controls~\citep{Montesino:2011kb,Aguirre:2012dz}. A more differentiated view that includes the automation of required security management processes as well is presented by \cite{Tracy:2007jq} where challenges, approaches and potential rewards of a security process automation platforms are discussed. The consideration of risk-aware business processes, their management and their actual implementation is the focus of work by \cite{Conforti:2013kw}. In our own work we propose a tool-supported continuous \ac{isms} approach~\citep{Brunner:2019fb} and empirically evaluated its  potential to automate \ac{isms} activities~\citep{Brunner:2017by} and means to introduce it in actual enterprise settings~\citep{Brunner:2018wx} via multiple case studies. 

\subsection{\acl{isrm}}

\begin{figure*}
\begin{center}
\includegraphics[width=.9\textwidth]{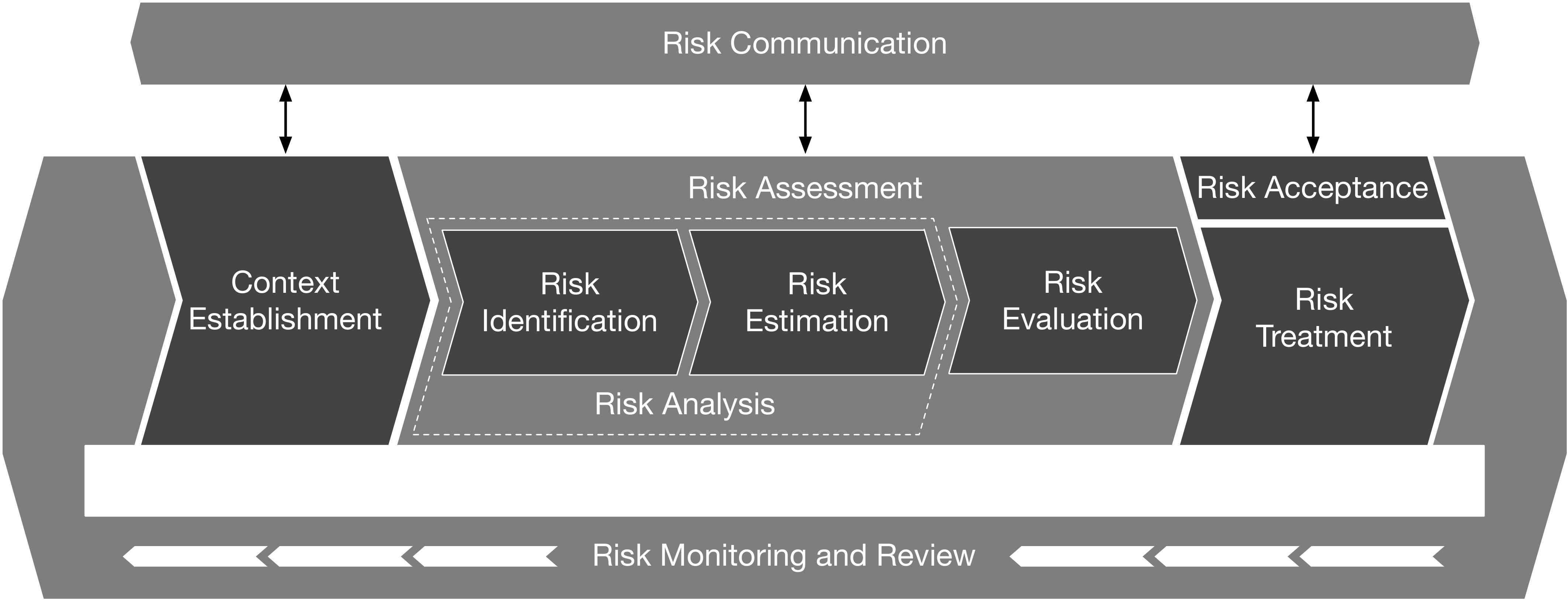}
\caption{\ac{isrm} process according to ISO 27005}
\label{fig:isrm-process}
\end{center}
\end{figure*}

Information security risk is defined as \shortquote{potential that a given threat will exploit vulnerabilities of an asset or group of assets and thereby cause harm to the organization} \citep{ISO:2011wd}. A general approach to systematically manage information security risks is outlined in the ISO 27005 standard \citep{ISO:2011wd}. The same basic building blocks and processes can be universally identified in nearly all \ac{isrm} standards, best practices and many other information security frameworks or risk management approaches. In contrast to many other risk management applications, the actual area of investigation is not restricted to singular domains (e.g., software development, IT service operation) or single projects, but most often covers the whole enterprise or a substantial part of it as strategically defined by the \ac{isms} scope. Figure~\ref{fig:isrm-process} illustrates this process which generally consists of 5 different activities where (1) context establishment, is followed by (2) risk assessment, with (3) risk treatment (and possibly acceptance of residual risks) concluding each iteration. Risks are further continuously (4) monitored and reviewed and (5) communicated within organizations. 

The first step, \emph{context establishment}, primarily deals with the definition of the scope and boundaries of the risk management initiative as well as providing the organizational principles to conduct \ac{isrm}. In context of an \ac{isms} this step will be aligned with the respective activities already mandated by the applied \ac{isms} standard. \emph{Risk assessment} includes the identification of risks, their estimation and evaluation. Risk identification will typically involve the documentation of relevant assets (i.e. within the scope), the identification of potential threats to and vulnerabilities of these assets. Taking established security controls into account potential consequences can then be identified resulting in a list of incident scenarios that might harm an organization's information security goals. These risk scenarios require subsequent estimation to assess their actual level of risk where enterprises can choose from a variety of qualitative (subjective and scale-based, e.g., critical, high, medium, low) and quantitative (calculated, e.g., expected annual financial loss) risk estimation approaches. This step will commonly include the determination of each risk's consequence and its respective likelihood. Depending on an organization's risk acceptance criteria, all information security risks above the agreed threshold will require proper \emph{risk treatment}. If possible and financially reasonable organization will define additional controls to reduce these risks to finally reach a satisfactory residual risk level. The \emph{monitoring} of information security risks and \emph{review} of security controls until the next cycle concludes each risk management iteration. The continuous \emph{communication} of information security risks and all information obtained from risk management activities is of crucial importance to ensure timely coordination between involved stakeholders. 

Established general-purpose \ac{isrm} standards include ISO 27005 \citep{ISO:2011wd}, the NIST SP800-30 Guidelines \citep{Stoneburner:2002tn} and the RiskIT Framework \citep{ISACA:2009ri}. Other management standards with a heavy emphasis on \ac{isrm} are the BSI IT Baseline Protection Methodology \citep{bsi:2017ms}, COBIT \citep{cobit5}, \ac{itil} \citep{Long:2011it} and the Common Criteria for Information Security Evaluation \citep{commoncrit}. Additional domain-specific standards cover \ac{isrm} practices, although within a more limited scope and typically present a specialization of a general-purpose standard for a given domain. 

Research has produced a variety of \ac{isrm} techniques. The more established ones with information security focus are ISRAM \citep{Karabacak:2005bh}, CRAMM \citep{Yazar:2002tq}, OCTAVE \citep{Panda:2009vo}, CORAS \citep{Lund:2011ip} and UMLsec \citep{Jurjens:2002jq} with the latter ones emphasizing formal model-driven approaches. They require a greater effort to build and maintain adequate system and enterprise models, but simultaneously offer features for formal security analysis. More general-purpose techniques, primarily considered in risk assessment, are \ac{fmea}, \ac{pha}, \ac{fta} or \ac{hazop} which all heavily rely on stakeholder expertise and have less stringent documentation and modeling requirements. 

Automation in \ac{isrm} is researched from different perspectives. An approach to automatically identify adequate security requirements based on an asset model of the system under investigation has been presented by \cite{Pasquale:2016dj}. Adaptive \ac{isrm} approaches enhance support for dealing with changes of assets as well the threat landscape \citep{Bennaceur:2014eha}. Automated risk analysis techniques using data flow analysis in business process models has been proposed by \cite{Accorsi:2012uf} and \cite{Berger:2016vv}. A tool-based approach automating threat analysis has been presented by \cite{Schaad:2012gk} but is strictly limited to asses software architectures. Due to the specialized nature of these approaches (either with regard to documentation effort, required stakeholder expertise or general applicability), none of them could have been directly evaluated in context of an organization's much more expansive \ac{isms}. 

\subsection{Conceptual Model}\label{sec:background:model}

\begin{figure*}
\begin{center}
\includegraphics[width=.9\textwidth]{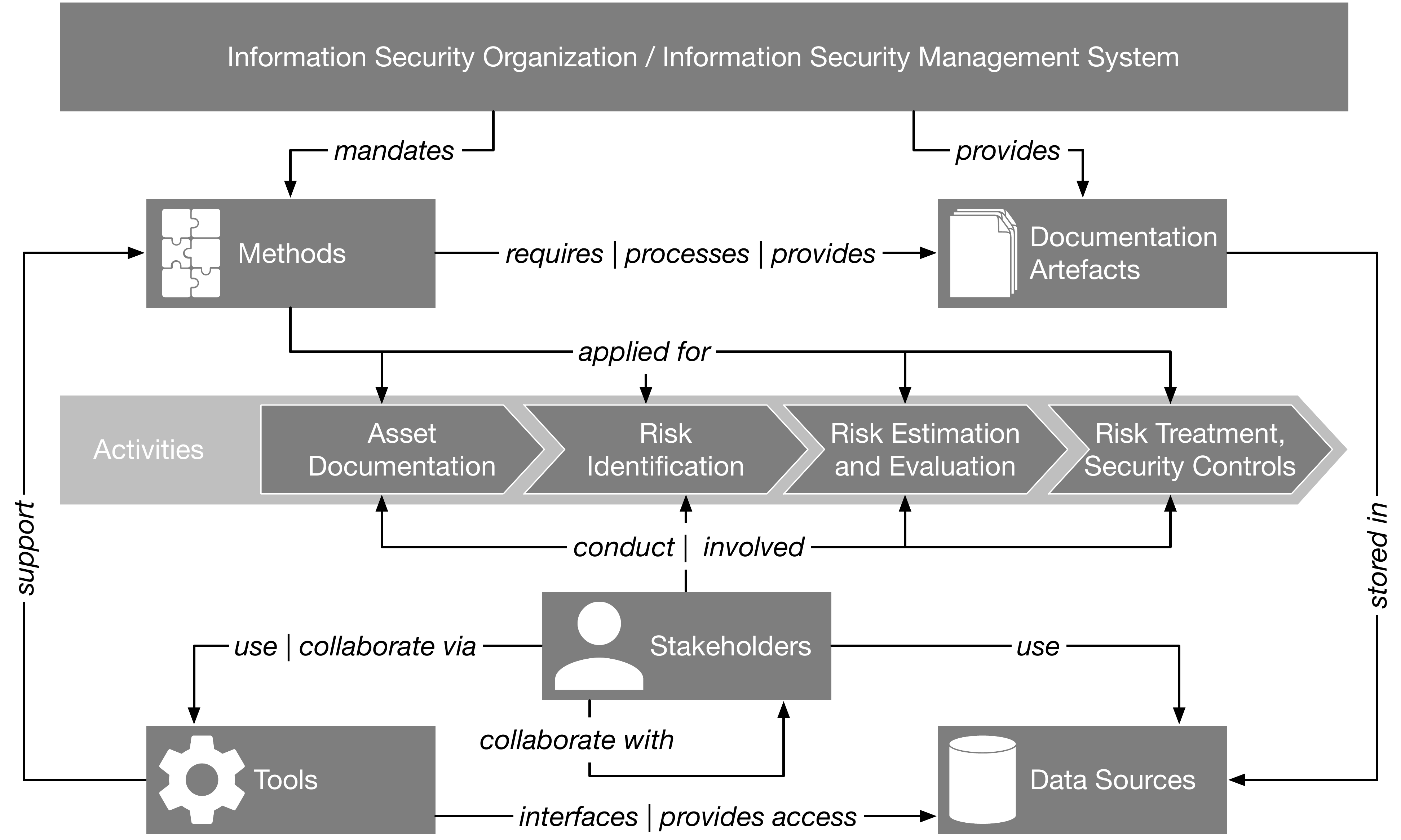}
\caption{Conceptual model for \ac{isrm} practices used in \ac{ism}}
\label{fig:conceptual-model}
\end{center}
\end{figure*}

The goal of our study was the determination of the status quo concerning risk management practices in \ac{ism}. As such we need to investigate the different approaches applied by enterprises to tackle risk management activities from multiple viewpoints. We thus developed a model from multiple \ac{isrm} and \ac{isms} standards and scientific publications. In addition we consulted multiple well-received practitioner guidelines in the information security management domain such as \cite{Disterer:2013is,Schou:2015is,Calder:2012gv,Watkins:2013in} and more. The conceptual model shown in Figure~\ref{fig:conceptual-model} illustrates how we conceptualize major activities, the relevant application environment and relations in between them.

The established \ac{isms} or information security organization defines the scope of all conducted risk management activities and will mandate methods to be applied in each step. Each method will require certain documentation artifacts which, in our conceptual model, are any formally or informally documented collections of information required for any given risk management activity. This includes high-level information security management policies (e.g., defining the scope and strategic alignment of an organization's information security activities) provided by the overarching \ac{isms} as well as all information being collected, provided, and potentially enriched by applied risk management methods. To gather a comprehensive picture we further distinguish between application of methods in distinct risk management activities. Expected documentation artifacts for asset documentation will include hardware inventories, process documentation or network plans, whereas risk identification will additionally deal with threats and vulnerabilities and provide documentation artifacts regarding risk scenarios. These will be enriched with a risk rating via risk estimation methods. Risk treatment will finally provide security controls as means to mitigate unacceptably high risks in accordance with an organization's risk treatment plan.

Different sets of stakeholders conduct risk management activities and utilize tools and datasources to complete their respective tasks. Tools will support one or more tasks and can be used as direct means to perform them (e.g., vulnerability scanner to identify vulnerabilities of an asset) or to orchestrate stakeholder collaboration (e.g., wiki to discuss individual risk ratings and reach a conclusive decision). Various data sources can be either directly used by stakeholders (e.g., vendor-specific security advisories) or interfaced by dedicated tools (e.g., vulnerability scanner integrates vendor-specific vulnerability database). Furthermore, different data sources can be used to store relevant documentation artifacts.

By investigating the concepts in this model -- the characteristics of each element and their relationships in actual organizational settings -- we can develop an encompassing picture of the current state of \ac{isrm} practices in \ac{ism} and identify potential challenges to be addressed in future research.

\section{Related Work}\label{sec:relatedwork}

Several studies have investigated the state of practice regarding \ac{ism} and \ac{isrm}. Investigations exploring the status quo on a broader setting are limited. Commonly research is restricted to individual aspects or individual risk management practices. In the following, we will thus discuss studies that we deem closely related to our research endeavor. These empirical studies include exploratory investigations, quantitative surveys, qualitative expert interviews, industrial case studies and systematic literature reviews. However, not all of them might necessarily cover the whole area of \ac{ism} or all aspects in \ac{isrm}.

The very stakeholder-centric point-of-view regarding information security policies, their implementation, perception and impact has been empirically and exploratory investigated in multiple articles. A study analyzing the differences in stakeholder perception regarding security policies has been conducted by \cite{Samonas:2020jl}. Compliance with and employees adherence to information security policies was investigated by \cite{Ifinedo:2014bb,Ifinedo:2012in} and \cite{Sikolia:2016tc}. Sommestad et al. conducted two noteworthy studies starting with the identification of variables influencing compliance with information security policies of organizations \citep{Sommestad:2014fs} and a further investigation assessing the theory of planned behavior to explain policy compliance \citep{Sommestad:2015br}. All listed articles try to shed light into employees willingness to follow information security policies and how to increase their adoption in organizational settings. Theoretical models that have been applied include the \ac{tpb} and \ac{pmt}. Apart from \cite{Sikolia:2016tc} most study results have been derived from smaller investigations either covering a singular organization or surveys with small sets of participants. Generalizability of findings has to be considered in need of improvement with studies pointing to heavy influences from distinct organizational settings. Nearly all studies conclude that top-management commitment is one of the most important drivers of information security compliance in organizations.

Research regarding stakeholder participation in \ac{isrm} practices and its influence in context of regulatory compliance has been presented by \cite{Spears:2010ud}. \cite{Rees:2008hv} and \cite{Montesdioca:2015jg} conducted surveys to investigate the user satisfaction with information security  and risk management practices. Other exploratory studies investigating the role of stakeholder knowledge for assessing data quality of documentation artifacts used in \ac{isrm} or for the expanded endeavor of \ac{grc} activities have been conducted by \cite{Sillaber:2015bv,Sillaber:2019gd}. These studies primarily investigate the satisfaction of various stakeholders with specific \ac{isrm} practices or their respective results. They do not provide a substantial investigation of these practices themselves, how they are implemented, which tools are used or which collaboration patterns are applied.

Moreover, a few empirical studies by Sauerwein et al. analyze the use of external and internal security information sources for \ac{isrm} processes. For example, \cite{Sauerwein:2019an} provides a comprehensive analysis of these information security sources used in research and practice. Most of these sources are unstructured and used in an ad-hoc manner by employees as input for critical information security and risk management processes without formal approval. \cite{Sauerwein:2018sh} empirically investigates this phenomenon and describes it as shadow threat intelligence.

Academic studies concerning the overall state of practice in \acp{isms} are sparse. The influence of organizational culture on \ac{isms} efficiency was analyzed by \cite{Chang:2007iwa}. The quality of information security management and implemented controls has been the focus of \cite{Baker:2007kj}. Several case studies focusing on  subtopics of ISRM from an organizational perspective can be identified. These case studies include investigations on how organizations conduct information security assessments based on standards~\citep{Shedden:2006risk}, how security risk assessment methods can more efficiently identify and treat the knowledge associated with business processes~\citep{Shedden:2011in} and the potential for improved asset identification enabled by the Rich Description Method (RDM)~\citep{Shedden:2016as}. A case-study based comparison of \acp{isms} has been conducted by \cite{vanWessel:2011gy} by investigating their implementation and impact in European and Chinese enterprises. \cite{Barlette:2010ad} conducted research into general international \ac{isms} adoption and derived corresponding drivers and success factors. These studies at most consider the \ac{isms} standard used by organizations and do not distinguish between differences in their implementation. Overall, we could not identify reliable research regarding actual \ac{isms} implementations, tool utilization and orchestration of stakeholder collaboration. In addition, many of other available studies such as \cite{Fitzgerald:2007ve,Pierce:2008ol,Hooper:2016em} primarily focus on the \ac{isms} top-management perspective. They exclusively involve roles such as \ac{ciso} or \ac{cio} and capture only parts of the \ac{isms} and \ac{isrm} activities under direct supervision of these subjects. 

Several researchers conducted empirical or literature studies to identify issues and challenges in \ac{ism} and \ac{isrm}. In this context, \cite{Fenz:2014cur} outline current challenges in \ac{isrm}, \cite{Wangen:2013wb} introduce a taxonomy of challenges in \ac{isrm}, \cite{Wangen:2016in} documents several issues concerning the application of qualitative and quantitative methods in \ac{isrm} practice, \cite{Soomro:2016in} argue the need of a more holistic approach for \ac{ism} and \cite{Webb:2014hg}  highlight deficiencies in the practice of information security risk assessment that lead to poor decision making and inadequate security strategies.

Looking beyond empirical studies performed in academic contexts, we identified several related whitepapers and reports published by IT and management consulting companies such as \cite{PWC:2015vq}, \cite{Deloitte:2017vt}, \cite{EY:2018gi} or \cite{MS:2019ps}. Other viable resources are available from common interest groups such as \cite{NISPlatform:2015wk} or \cite{SANS:2019cs}. Considering obvious financial motives behind most of these studies, their conclusions should be critically questioned. However, certain reoccurring subjects in these studies are notable and corroborate academic findings such as challenges in reliably evaluating an organization's risk exposure and the alignment of business and technical perspectives. The continuous evaluation of systems within information security management and risk assessment is another open and reoccurring challenge that -- in light of more flexible supply chains, increase in usage of distributed services and  an overall increase in information system complexity -- requires further attention.

The available body of academic (and also non-academic) research is inconclusive to fully address our research objectives stated in Section~\ref{sec:introduction}. We thus conducted a thorough empirical and exploratory investigation into the current state of practice regarding \ac{isrm} practices in \ac{ism}. Contrary to available research, a broader study that emphasizes not just the strategic and top-management roles' perspective but that draws a conclusive picture of actually implemented risk management practices including used methods, tools, documentation artifacts and collaboration patterns is required to better guide future research in this domain.

\section{Research Method}\label{sec:method}

We used an anonymous online survey instrument for the purpose of this exploratory investigation and followed respected guidelines for the design and execution of our research~\citep{Kitchenham:2003ps,Kasunic:2005vt} and proactively addressed challenges in survey research~\citep{Wagner:2019ui}. A pilot questionnaire and a subsequent interview with an experienced \ac{isms} manager were conducted to ensure the validity and content as well as the general usability of the developed survey instrument. Received feedback allowed us to make a few minor changes to some multiple choice answer options. In this section, we present the final results of our study design process. 

\subsection{Research Questions}\label{sec:method:rqs}

Our primary research goal was to gain insight into the current state of risk management practices in \ac{ism} and to use this information to further identify practical and directly applicable means for improvement. Considering the sensitive topic at hand we refrained from directly asking study participants to reveal the inner workings of their \ac{isms} and especially their risk management practices. In addition, the expected heterogeneous nature of these processes would have further complicated the design of an efficient online survey. 

Instead, we developed an approach to collect vital information regarding our research objectives which will not require study participants to disclose sensitive information. In accordance to our conceptual model (cf. Section~\ref{sec:background:model}) we set the focal point of our investigation on the considered artifacts, involved roles, methods and tools which further proved to be a more reliable way to collect the desired data during initial pilot interviews. We consecutively derived the following research questions for our survey:
\begin{description}
	\item[RQ1] What methods and documentation artifacts are considered for risk analysis in ISMS?
	\item[RQ2] Which stakeholders are involved in the risk analysis and how do they cooperate? 
	\item[RQ3] What information sources and tools are used for risk analysis in ISMS?
\end{description}
By investigating these research questions we can gain a better understanding of the current state of practice and identify potential areas for improvement with regard to stakeholder collaboration, artifact documentation and tool usage in \ac{isrm} activities. 

\subsection{Instrument}\label{sec:method:survey}

We structured our online questionnaire in six question blocks. These cover \emph{demographics} (DE), general \emph{information security management} practices (IM), \emph{asset documentation} (AS), \emph{risk identification} (RI), \emph{risk estimation and evaluation} (RE) and the documentation of \emph{security requirement and controls} (SE). Overall, the questionnaire contains 45 questions and was aimed at 20 minutes duration for completion.

Table~\ref{tab:survey-questions} shows an excerpt of the designed survey, the complete survey instrument is provided in Appendix~\ref{app:survey}. The table's first column contains the question ID which is superseded by a question block indicator. The second column contains the question itself. The last column shows the type of question: \emph{Single} choice, \emph{Multiple} choice, \emph{Yes/No}, \emph{Numeric}, \emph{Rating}, \emph{Ranking} and \emph{Open}. Ratings are used to inquiry specific implementation aspects  of different \ac{isrm} tasks (e.g., "Risk identification is performed automatically."). We used an ordinal scale of 4 ("Applies fully", "Applies mostly", "Applies to some extend", "Does not apply") with the additional option "Do not know" to capture when participants were not knowledgeable. Rankings were used to capture the importance of utilized tools following multiple-choice questions where more than 5 options were provided (cf. questions RI003 and RI004 in Table~\ref{tab:survey-questions}). If a single or multiple choice question allowed participants to extend answer options via a dedicated text input it is marked with~\textsuperscript{+} and if participant were able to provide an additional comment to a question this marked with~\textsuperscript{*} respectively. Both options were used to allow participants to express company-specific deviations and additions.

\begin{table*}
\caption{Survey Questions (excerpt)}
\begin{center}
\begin{tabularx}{\textwidth}{lXl}
\toprule
\textbf{Id} & \textbf{Question} & \textbf{Type} \\
\toprule
DE001 & What is your organizational role?& Multiple\textsuperscript{+}\\
DE002 & Which of the following personal certifications and qualifications do you have? & Multiple\textsuperscript{+}\\
DE003 & How many years of professional experience in the area of information security do you have? & Numeric\\
DE004 & What type of industry is your organization in? & Single\textsuperscript{+}\\
DE005 & What is the size of your organization? & Single\\[-0.5em]
\ldots & \ldots & \ldots\\
IM001 & Do you operate an Information Security Management System (ISMS)? & Single\textsuperscript{*}\\
IM002 & What is the most important driver for Information Security Management in your organization? & Multiple\textsuperscript{+}\\
IM003 & Which Information Security Risk Management (ISRM) methodology does your organization apply? & Multiple\textsuperscript{+}\\
IM004 & How often does your organization conduct an information security risk management cycle? & Multiple\textsuperscript{+}\\
IM005 & Which events additionally trigger information security risk management activities in your organization? & Multiple\textsuperscript{+}\\
IM002\textsubscript{alt} & What do you consider the most important driver for Information Security Management? & Multiple\textsuperscript{+}\\
IM003\textsubscript{alt} & Which Information Security Risk Management (ISRM) methodology do you know? & Multiple\textsuperscript{+}\\[-0.5em]
\ldots & \ldots & \ldots\\
RI000 & Does your organization perform Information Security Risk Management (ISRM) or related activities? & Yes/No\\
RI001 & Rate the following statements with regard to the information security risk identification approach of your organization. & Rating \\
\hspace{.25cm} Stmt1 & Risk identification is performed automatically. & \\
\hspace{.25cm} Stmt2 & Every relevant security risk is identified in a timely fashion. & \\
\hspace{.25cm} Stmt3 & Sharing of relevant security information is conducted via a formal process. & \\
\hspace{.25cm} Stmt4 & Relevant security information is automatically preprocessed and filtered for conducting risk identification. & \\
\hspace{.25cm} Stmt5 & Exchange of security information with other organizations and individuals has been beneficial for risk identification. & \\
RI002 & Which aspects are considered for information security risk identification in your organization? & Multiple\textsuperscript{+}\\
RI003 & Which EXTERNAL information sources are used for information security risk identification in your organization? & Multiple\textsuperscript{+}\\
RI004 & Which are the three most important EXTERNAL information sources for information security risk identification in your organization? & Ranking\\[-0.5em]
\ldots & \ldots & \ldots\\
 RI011 & What are the most pressing challenges during information security risk identification for your organization? & Open\\[-0.5em]
\ldots & \ldots & \ldots\\
\bottomrule
\end{tabularx}
\end{center}
\label{tab:survey-questions}
\vspace{-1em}
\end{table*}%

Due to natural differences between enterprises already operating an \ac{isms}, and those still in the planning phase, as well as participant involvement in and knowledge of dedicated \ac{ism} activities we implemented alternative paths in our survey. Questions with subscript "alt" were only shown to participants which answered previous questions to the negative (e.g., IM002\textsubscript{alt} and IM003\textsubscript{alt} were shown to participant not operating an \ac{isms} instead of questions IM002 to IM005).

Each question block was further complemented by descriptive text providing additional context for every question. Terms and definitions used throughout the questionnaire and within this descriptive blocks were taken from the ISO 27k family of standards~\citep{ISO:2013wu}. The description for the risk identification block is provided as example below:

\begin{quote}
This group of questions asks specifics about the Information Security Risk Management (ISRM) approach at your organization. In particular these questions target the way how risks are identified at your organization as part of your ISMS or ISRM initiative. According to ISO 27005 Risk Identification is "[...] the process to find, list and characterize elements of risk".
\end{quote}

Apart from the question block for demographics, each block contains dedicated questions directly linked to our research questions. We ask participants for applied methods and characteristics of documentation artifacts (RQ1), involved stakeholders as well as their mode of collaboration (RQ2) and used data sources and tools (RQ3). 

\subsection{Data Collection Procedure}\label{sec:method:collection}

Our target audience were companies and organizations which had already established an \ac{isms} or were planning to do so in the near future. We explicitly did not focus on companies which already operated a certified \ac{isms} since this would greatly decrease our potential survey population without necessarily raising the result quality for this exploratory investigation. Targeted participants were  stakeholders being responsible or involved in strategic or operational \ac{ism} activities at these companies. Again, we did not exclusively target top management information security roles (e.g., CISO, CIO) as this would have unnecessarily restricted our population and would potentially lead to a more strategic point of view instead of the targeted practical one. Furthermore, top level \ac{ism} roles are not guaranteed to be in place at targeted companies, especially the ones who are currently introducing or only planning to implement an \ac{isms}. 

Participants were acquired through multiple channels. We invited participants from Austrian research transfer projects Qualifizierungsnetzwerk-West (Q-West), Digital Tourism Experts (DTE) and Digitalisierung und Sicherheit (DuS). These were prime candidates as they attended practical information security management courses covering various risk management practices with the clear intention to either introduce systematic \ac{ism} in their organizations or improve their current \ac{isrm} practices. In addition we invited German information security experts through a dedicated security common interest group's mailing list. Finally, we directly contacted and invited respected \ac{ism} and \ac{isrm} experts from industry. The data collection phase for these participant groups diverged and is shown in the data collection overview in Table~\ref{tab:participant-sources}. In total, 351 participants were invited between November 2017 and July 2019. Invitations were delivered via e-mail and all participants were informed upfront that responses were collected anonymously. The used invitation letter is provided in Appendix~\ref{app:invitation}. We used the same self-hosted and administrated survey tool instance (Lime Survey -- \url{https://www.limesurvey.org}) for all participant groups.

\begin{table}
\caption{Data collection (overview)}
\begin{center}
\begin{tabularx}{\linewidth}{Xccc}
\toprule
\textbf{Group} & \textbf{Country} & \textbf{Invited} & \textbf{Data collection phase}\\
\toprule
Q-West & Austria & 34 & 2017-11 -- 2018-01 \\
DTE & Austria & 43 & 2018-05 -- 2018-10 \\
DuS & Austria & 46 & 2019-07 -- 2019-09 \\
\midrule
Security Interest Group & Germany &  \textasciitilde200 & 2018-05 -- 2019-09 \\
\midrule
Direct & DACH & 28 & 2017-11 -- 2018-12 \\
\bottomrule
\end{tabularx}
\end{center}
\label{tab:participant-sources}
\end{table}%

\subsection{Data Analysis Procedure}\label{sec:method:analysis}

Data analysis was, due to the use of an online questionnaire and export capabilities of the survey tool, straight forward. Responses were exported and briefly checked for completeness and consistency. Obviously incomplete responses were removed. That concerned responses where less than $66\%$ of questions were answered or participants made excessive use of ignoring all non-mandatory questions. Furthermore all responses were discarded where participants showed no knowledge of any concerned \ac{isrm} activity (e.g., answering "I do not know" for all statement ratings in multiple question blocks like RI001 shown in Table~\ref{tab:survey-questions}). 

We then conducted a qualitative analysis of the remaining responses, created descriptive statistics and examined response patterns. We prepared appropriate graphical presentations. The data was summarized and reported for all questions in the survey to address our research questions. Table~\ref{tab:questions-rqs} shows the mapping between our research and survey questions. Finally, we applied manual blocking to our results. Our main area of interest were differences between enterprises operating an ISMS and those planning to do so.

\begin{table}
\caption{Mapping between research and survey questions}
\begin{center}
\begin{tabularx}{\linewidth}{lp{3.35cm}X}
\toprule
 & \textbf{Research Question} & \textbf{Survey Questions} \\
\toprule
RQ1 & What methods and documentation artifacts are considered for risk analysis in ISMS? &  IM001, IM002, IM003, IM004, IM005, AS001, AS002, RI001, RI002, RI010, RE004, RE005, RE006, RE007, RE009, SE001, SE006 \\
RQ2 & Which stakeholders are involved in the risk analysis and how do they cooperate? & RI007, RI008, RI009, RE001, RE002,  RE003, SE002, SE003 \\
RQ3 & What data sources and tools are used for risk analysis in ISMS? & AS003, RI003, RI004, RI005, RI006, SE004, SE005 \\
 --- & Challenges in respective areas  &  RI011, RE008, SE007\\
\bottomrule
\end{tabularx}
\end{center}
\label{tab:questions-rqs}
\end{table}%

\subsection{Validity Procedure}\label{sec:method:validity}

We performed several steps to check and ensure the validity of our research. Considering the often heterogeneous nature of implemented \ac{ism} and \ac{isrm} activities and the sometimes ambiguous use of terms, we used the generally accepted ISO~27005 standard~\citep{ISO:2011wd} as referential basis for our survey. All terms were clarified by additional descriptions for each question block and participant were given ample opportunity to state any issues within free-text comment blocks. We further performed a small pilot study and incorporated received feedback prior to the distribution of our questionnaire. Finally, during data analysis we planned to discard any results of questionable quality where participants made multiple obviously conflicting statements. This quality assurance step was primarily intended for responses where participants stated to operate a certified \ac{isms} or standardizes ISRM approach but would not perform basic mandatory activities, completely neglect required asset types, or resign to manage documentation artifacts demanded by the certified standard (e.g., operate an ISO 27001 compliant \ac{isms} but not document security controls, operating an \ac{isms} but not performing any kind of risk identification). 

\section{Results}\label{sec:results}

In this section, we present the survey results in relation with our research questions. We additionally address differences between companies with and without an established \ac{isms}.

\subsection{Study Population}\label{sec:results:population}

In total, we collected 64 responses of which 26 were processed for data analysis. The other 38 responses were dropped with the majority having aborted the survey within the first question block (cf. Section~\ref{sec:method:analysis}). No responses had to be dropped due to obviously conflicting statements (cf. Section~\ref{sec:method:validity}). Considering the amount of forwarded invitations we reached a response rate of $7\%$ which is comparable to other exploratory surveys (cf. Section~\ref{sec:relatedwork}) in this field. The remainder of this section presents the results derived from the 26 complete responses.

\subsubsection*{Participants}\label{sec:results:participants}
\begin{figure*}
\begin{center}
\includegraphics[width=\textwidth]{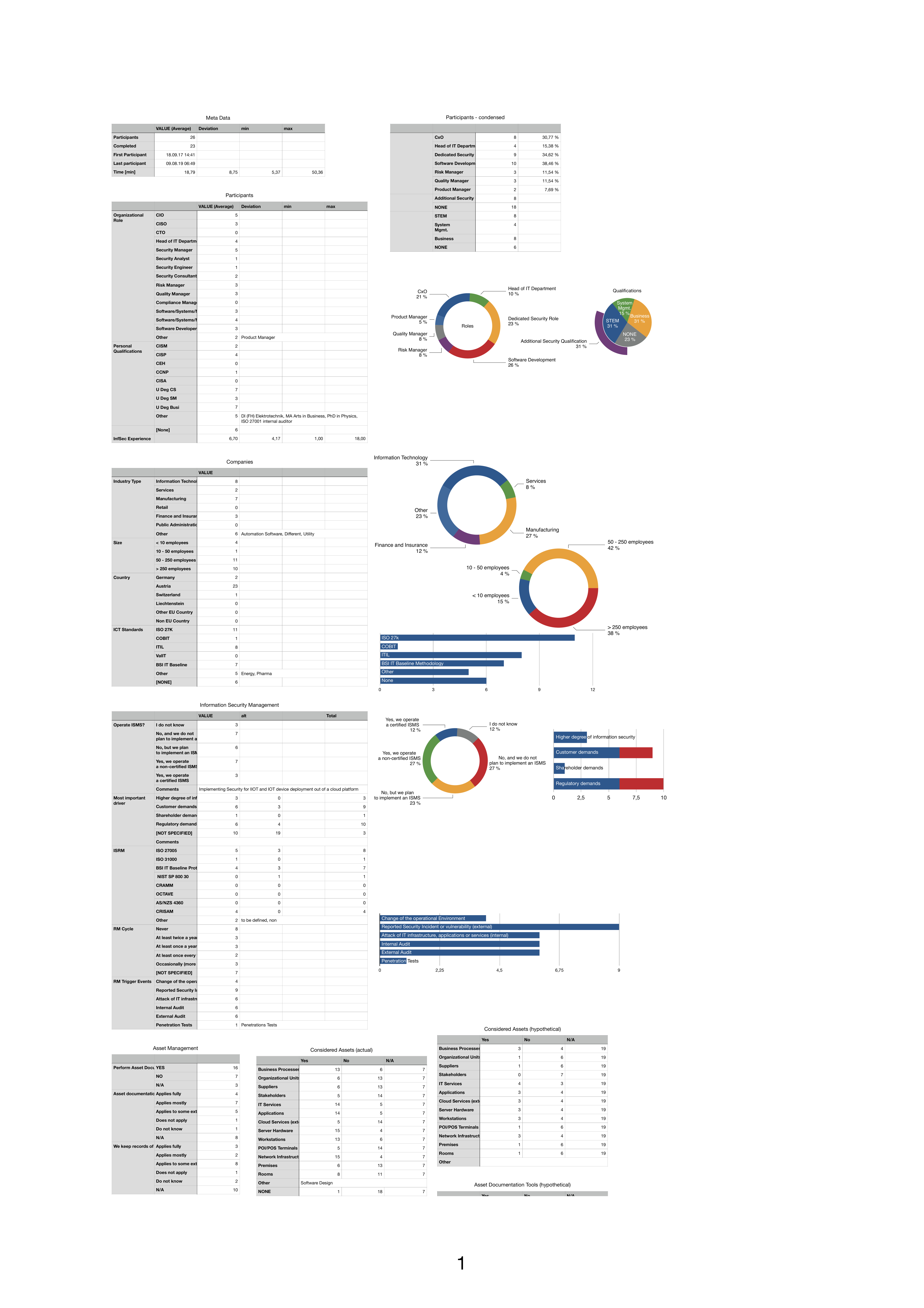}
\caption{Participant roles and qualifications ($N=26$)}
\label{fig:results-roles}
\end{center}
\end{figure*}

Of our participants, 8 gave \ac{ciso} or \ac{cio} as their organizational role with additional 4 participants being employed as head of IT department. This amounts to roughly one third of responses from higher-level management. Other participants mainly worked in dedicated security or risk management roles and in software development. Figure~\ref{fig:results-roles} shows the overall picture of participant roles as well as their qualifications. 20 participants had obtained a university degree in either science, technology, engineering and mathematics (STEM), system management or business programs. 8 participants had additionally obtained specialized personal security qualifications like \ac{cissp} or \ac{cism}. Participants stated an average of $6.7$ years of professional experience in information security (deviation: $4.17$, min: $1$ year, max $18$ years). While we cannot guarantee that multiple participants from one company contributed to our data set, the analyzed responses (and differences in answers) indicate that this was not the case.

\subsubsection*{Companies}\label{sec:results:companies}

\begin{table}
\caption{Business domains of responding companies ($N=26$)}
\begin{center}
\begin{tabularx}{\linewidth}{Xrl}
\toprule
\textbf{Business Domain} & \multicolumn{2}{l}{\textbf{Companies}} \\
\toprule
Information Technology & 9 & ($34\%$)\\
Manufacturing & 8 & ($31\%$)\\
Finance and Insurance & 3 & ($11\%$)\\
Services & 2 & ($8\%$)\\
Utilities & 2 & ($8\%$)\\
Healthcare & 1 & ($4\%$)\\
\midrule
\textit{not disclosed} & 1 & ($4\%$)\\
\bottomrule
\end{tabularx}
\end{center}
\label{tab:results-companies1}
\end{table}%

\begin{table}
\caption{Sizes of responding companies ($N=26$)}
\begin{center}
\begin{tabularx}{\linewidth}{Xrl}
\toprule
\textbf{Size} & \multicolumn{2}{l}{\textbf{Companies}} \\
\toprule
less than 10 employees & 4 & ($15\%$)\\
10 - 50 employees & 1 & ($4\%$)\\
50 - 250 employees & 11 & ($42\%$)\\
more than 250 employees & 10 & ($39\%$)\\
\bottomrule
\end{tabularx}
\end{center}
\label{tab:results-companies2}
\end{table}%
 
The most prominent business domains in our data set were \shortquote{Information Technology} (9 organizations, $34\%$) and \shortquote{Manufacturing} (8 organizations, $31\%$). Medium-sized (50-250 employees) and large (more than 250 employees) companies made up the majority of responses. Tables \ref{tab:results-companies1} and \ref{tab:results-companies2} show the results regarding business domain and company size. The majority of responses was provided from Austrian companies ($88\%$), although $20\%$ of them operate internationally with subsidiaries in at least one other EU or non-EU country as well.

\begin{figure}
\begin{center}
\includegraphics[width=.9\linewidth]{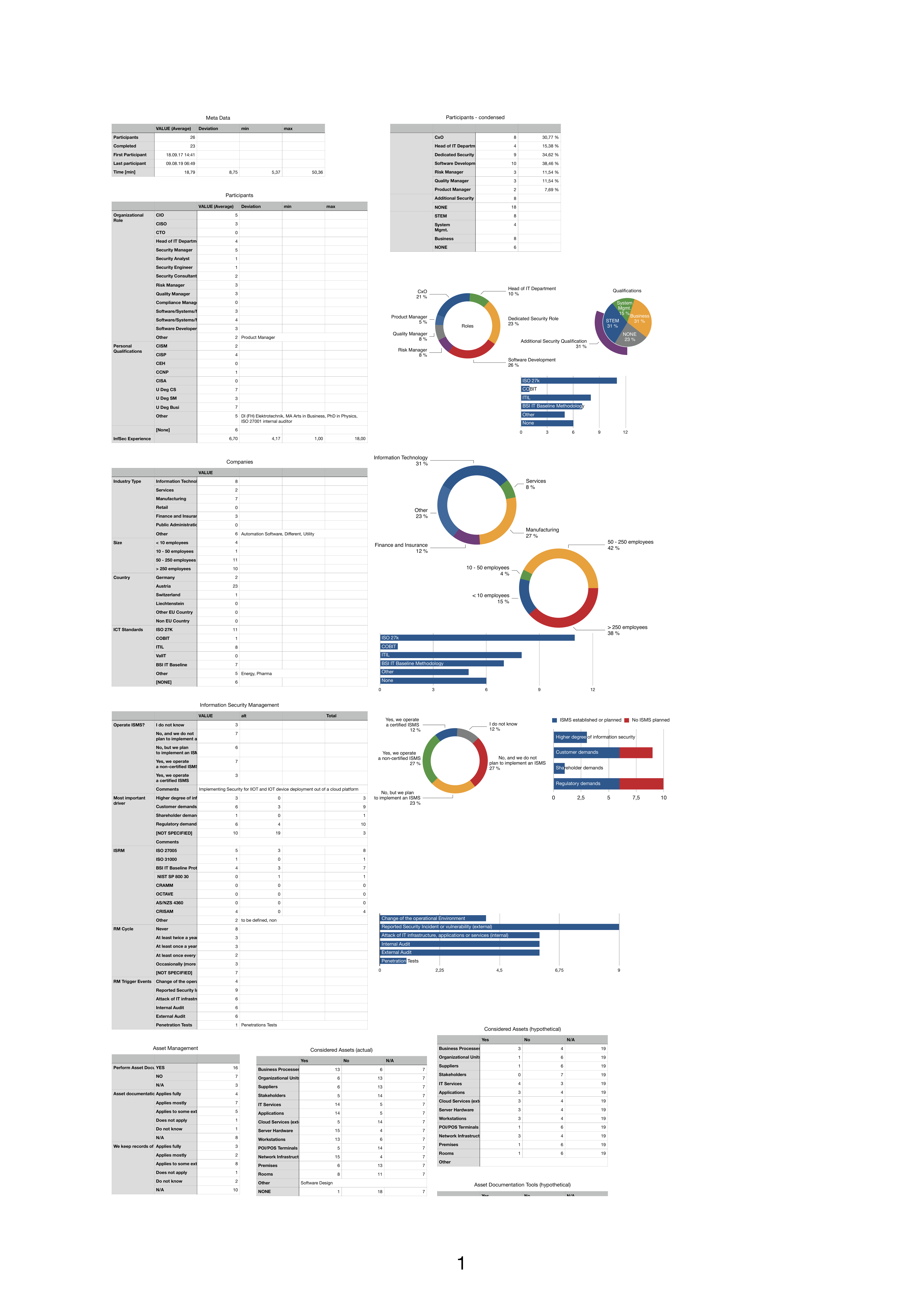}
\caption{ICT Standards being partially implemented ($N=26$)}
\label{fig:results-ict}
\end{center}
\end{figure}

When asked about implemented ICT standards, participants of 11 companies ($42\%$) stated that they at least partially implement some standard of the ISO 27k family with another 7 companies ($27\%$) using the BSI IT Baseline Protection Methodology or at least parts of it. COBIT is only used by one internationally operating, large enterprise. Standardized IT service management is a concern for 8 companies ($31\%$) as can be seen by the high adoption rate of the \ac{itil}. Other ICT standards directly named by participants concerned domain-specific requirements for utility or healthcare enterprises. Figure~\ref{fig:results-ict} illustrates these results.

\subsubsection*{Response Fragmentation}

\begin{figure*}
\begin{center}
\includegraphics[width=\textwidth]{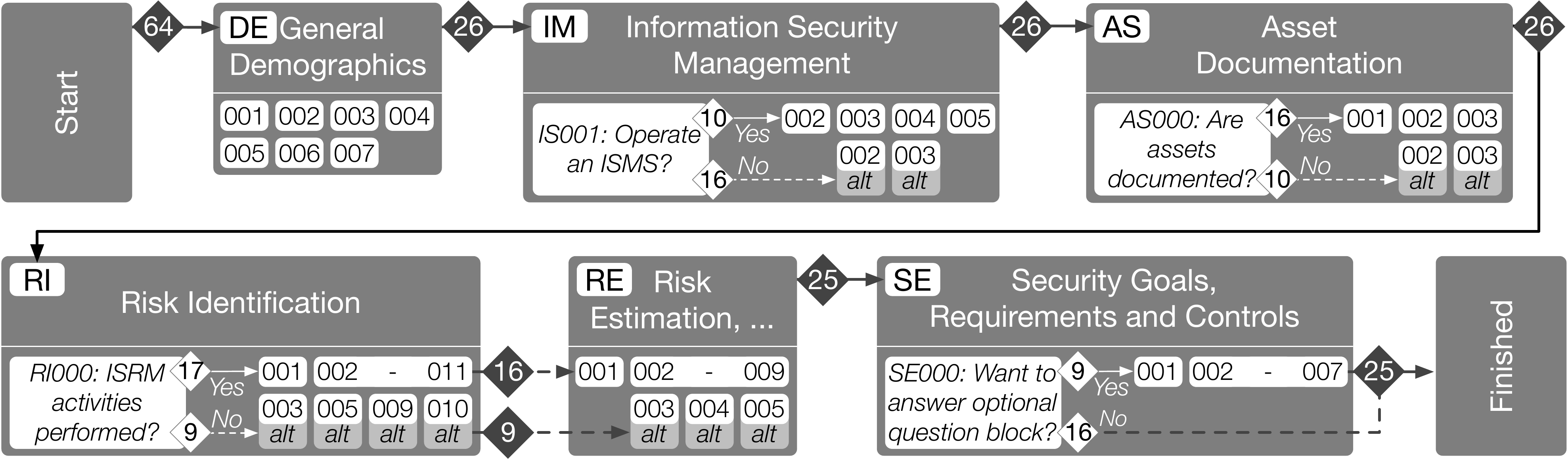}
\caption{Survey structure and number of responses for each query path}
\label{fig:results-fragmentation}
\end{center}
\end{figure*}

Our survey was designed in a way that participants were shown alternative questions if they did not implement specific \ac{isrm} activities in their organizations. The reasoning behind that step was that not every enterprise performing information security risk analysis would also necessarily operate an \ac{isms} or document their assets in a structured fashion. Furthermore, we wanted to capture the opinion of those who worked in organizations not yet operating an \ac{isms} as well. This design decision ultimately resulted in fragmented responses with different answer counts for individual question blocks. Figure~\ref{fig:results-fragmentation} shows the overall survey structure and the number of responses for each primary or alternative query path (numbers in diamond shapes at transitions between query blocks). Note that 38 respondents canceled the survey before completing the first question block and another respondent did so during the risk identification block. Thus, the reduced total number of responses between respective question blocks. Of the 26 responses 13 answered all questions in the main path of which seven in addition answered the optional block concerning security goals, requirements and controls.

\begin{figure*}
\begin{center}
\includegraphics[width=\textwidth]{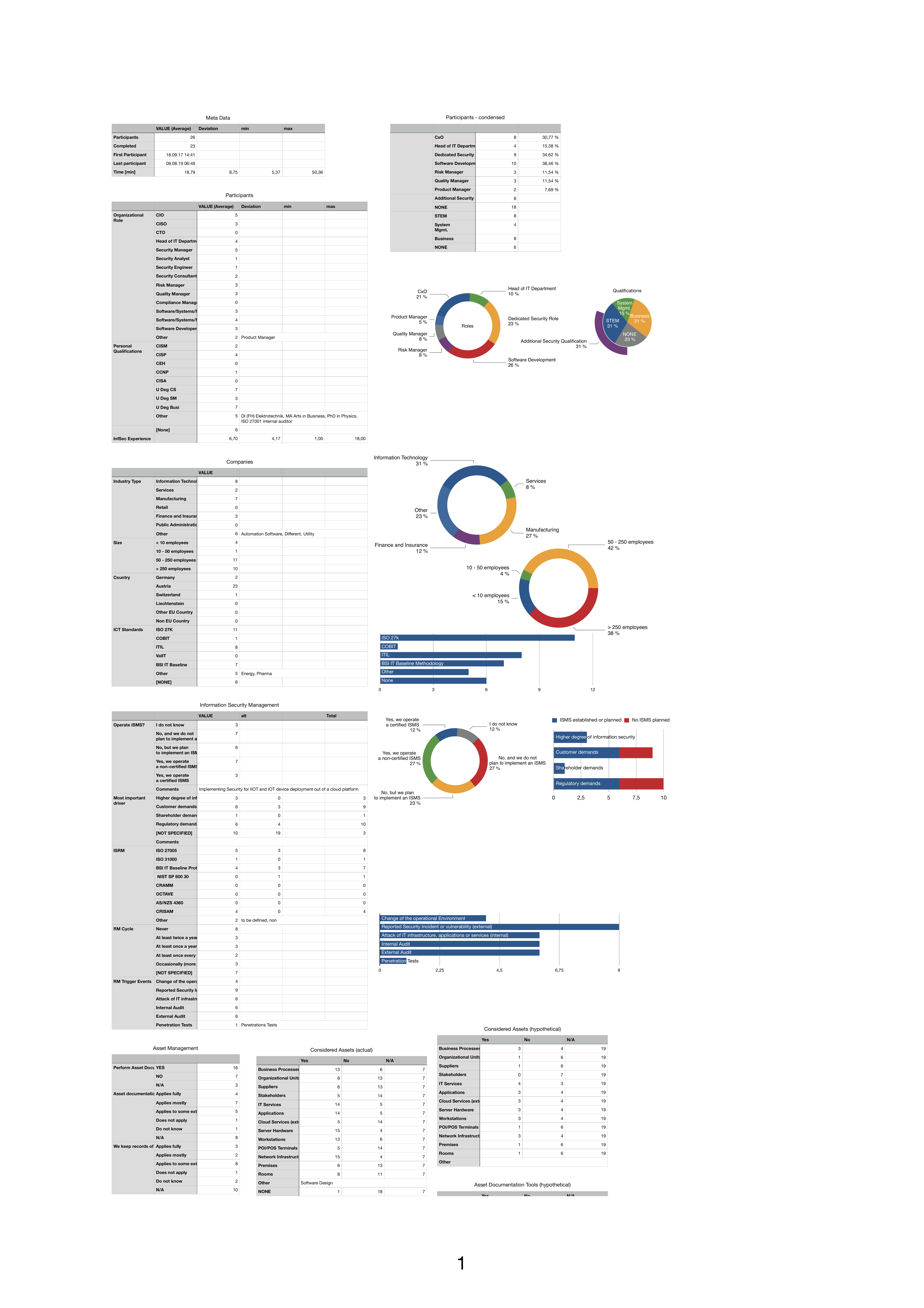}
\caption{\ac{isms} adoption and motivation ($N=26$)}
\label{fig:results-ism}
\end{center}
\end{figure*}

\subsection{\acl{isms} Adoption}\label{sec:results:ism}
Within our study population were 10 companies that have already established an \ac{isms} of which 3 achieved a corresponding certification. Another 6 companies were planning to implement an \ac{isms}. Figure~\ref{fig:results-ism} shows these results and the stated motivation for doing so. The two most common drivers for implementing an \ac{isms} were customer and regulatory demands, a view also shared by those who refrained from operating an \ac{isms}.

\subsection{Methods and Documentation Artifacts Considered for Risk Analysis in \ac{ism} (RQ1)}\label{sec:results:rq1}

\begin{figure}
\begin{center}
\includegraphics[width=\linewidth]{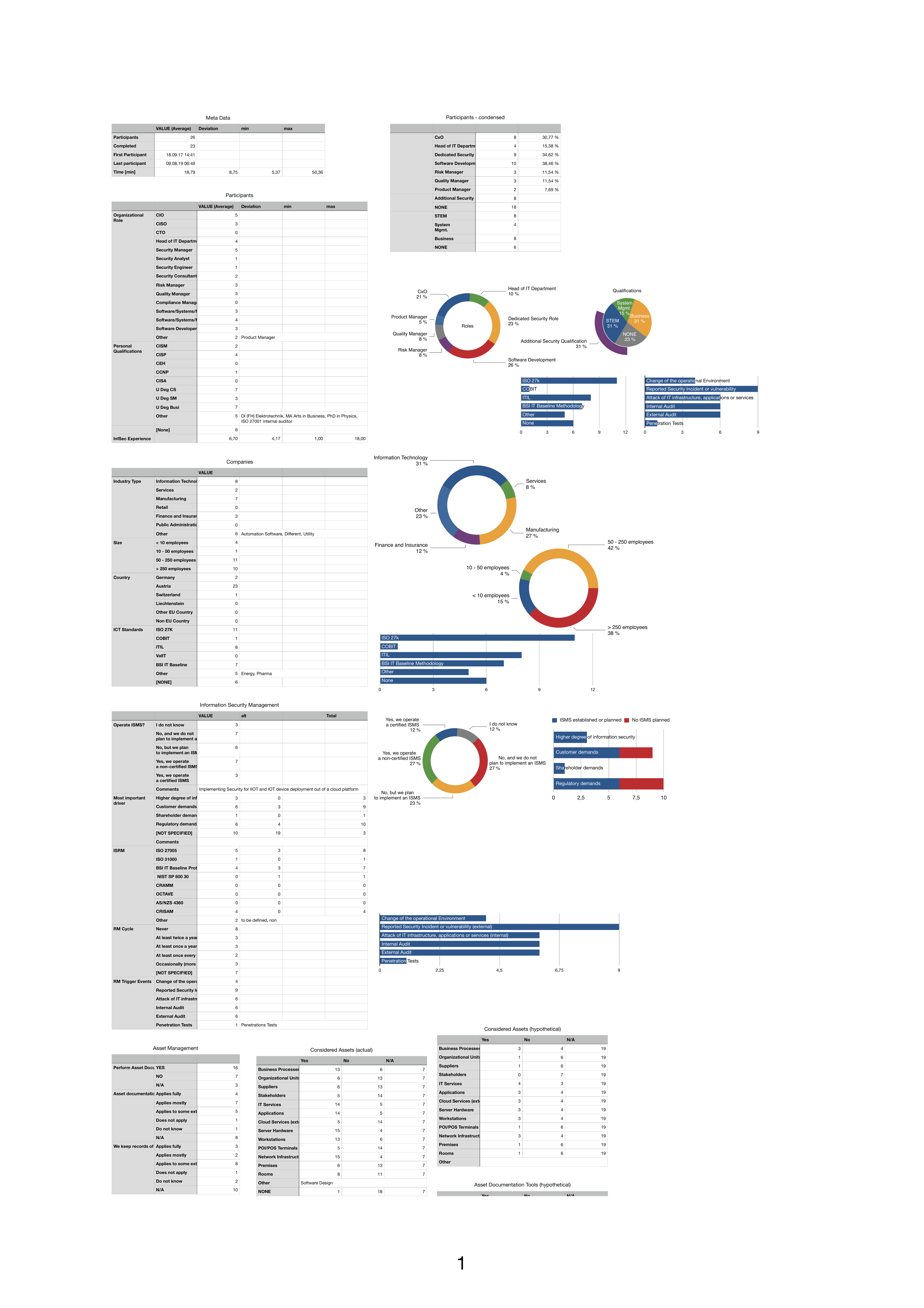}
\caption{Events triggering additional risk management cycles ($N=26$)}
\label{fig:results-rm-triggers}
\end{center}
\end{figure}

While not all participants disclosed the \ac{isrm} methodologies used in their organization, we identified ISO 27005 (5 companies), BSI IT Baseline Protection Methodology (4 companies) and CRISAM\footnote{CRISAM is a popular GRC toolsuite in the DACH region. It provides its own best-practice approaches for various \ac{isrm} tasks and in addition supports common information security standards.} (4 companies) to be the most commonly applied methods. Discipline regarding the regular execution of risk management cycles (as required by all of these standards) is not satisfactory. Although ten companies ($38\%$) had established an \ac{isms} and should thus regularly conduct risk management, only six ($23\%$) performed this activities at least on an annual basis. Half of the participating companies do not regularly perform risk management activities or with two or more years in between. Figure \ref{fig:results-rm-triggers} shows which events trigger additional risk management cycles. Reported security incidents or vulnerabilities are considered by almost every enterprise operating an \ac{isms} to warrant an unscheduled re-assessment of information security risks. Furthermore, actual attacks on IT infrastructure as well as internal or external audits commonly lead to additional risk management activities. Interestingly, only 4 enterprises stated that changes of the operational environment trigger a risk management cycle (cf. ISO 27001: \shortquote{The organization shall perform information security risk assessment at planned intervals or when significant changes are proposed to occur,[..]}). Regarding applied methods and corresponding documentation artifacts we identified a diverse overall picture. 

\subsubsection*{Asset Documentation}
Table~\ref{tab:results-artifacts} shows the types of assets that organizations conducting \ac{isrm} activities document and use. Note that only 16 organizations stated that they document assets at all. It can be observed that technical aspects (e.g., hardware, applications) are emphasized over organizational aspects (e.g., stakeholders, suppliers).

\begin{table}
\caption{Assets documented and used for information security risk assessment activities ($N=16$)}
\begin{center}
\begin{tabularx}{\linewidth}{Xc}
\toprule
\textbf{Types of Assets} & \textbf{Considered by}\\
\toprule
Network Infrastructure, Server hardware & $94\%$\\
IT Services, Applications & $88\%$ \\
Workstations, Business Processes & $81\%$ \\
Rooms & $50\%$ \\
Organizational Units, Suppliers, Premises & $38\%$ \\
Stakeholders, Cloud Services (external), POI/POS Terminals & $31\%$ \\
Other: Software Design & $6\%$ \\
\midrule
\textit{not disclosed} & $6\%$ \\
\bottomrule
\end{tabularx}
\end{center}
\label{tab:results-artifacts}
\end{table}%

Concerning the practical implementation of the asset documentation process, we could observe that this crucial activity is in most cases centralized ($69\%$ of responses stated that this activity is performed fully or mostly in a centralized manner). However, only $19\%$ of responses stated that they keep records of all individual assets, most others only keeping records of individual assets for certain groups. A similar picture can be observed for the timeliness of the available asset documentation, where only $19\%$ of the enterprises stated that their asset documentation is always up-to-date. This clearly relates to the still improvable automation of asset documentation and discovery ($88\%$ of companies stated that asset documentation is a mostly manual task) and the dominating practice of less-than-annual risk management cycles.

\subsubsection*{Risk Identification}

Information security risk identification is primarily performed in a manual fashion with only $31\%$ of organizations stating that they at least partially automated some of these activities. Only one organization identifies relevant security risks in a timely fashion, all others stating at least some deficiencies. In addition, a general lack of suitable internal processes to share relevant security information can be derived from the survey responses. Inter-organizational exchange of security information is generally not perceived as beneficial.

The most commonly applied risk identification methods are rather informal ones such as the use of checklists and brainstormings (cf. Table~\ref{tab:results-ri-methods}). More demanding approaches (with regard to methodological complexity, dedicated stakeholder expertise and documentation requirements) are sparingly used in context of \ac{ism}. Seven companies applied a mix of (2 to 5) risk identification technique compared to 5 companies relying on a singular method to identify information security risks. 

\begin{table}
\caption{Methods and practices applied for the identification of information security risks ($N=17$)}
\begin{center}
\begin{tabularx}{\linewidth}{Xc}
\toprule
\textbf{Risk Identification Methods} & \textbf{Applied by}\\
\toprule
Checklists & $53\%$\\
Brainstorming & $41\%$ \\
\ac{fmea}, CRISAM & $18\%$ \\
\ac{cca}, \ac{isram} & $12\%$ \\
\ac{swift}, \ac{pha}, \ac{ram} & $6\%$ \\
\midrule
\textit{not disclosed} & $12\%$\\
\bottomrule
\end{tabularx}
\end{center}
\label{tab:results-ri-methods}
\end{table}%

\subsubsection*{Risk Estimation, Evaluation and Treatment}

The most common approach to estimate identified information security risks is to rely on qualitative ratings of involved stakeholders ($46\%$ of responding companies, additional $23\%$ use semi-quantitative approaches). Only one organization (large internationally operating enterprise, using COBIT) performs quantitative risk estimation. Other responses did not disclose their risk estimation approach. Dependencies between assets are regularly considered during risk identification (only two companies ignore them) as are potential dependencies between security risks (ignored by three companies). Risk estimation on the other hand has a tendency to not incorporate estimated likelihood or probability of related risks. Furthermore, companies do not use automation techniques to perform risk estimation and solely rely on manual practices.

Regarding methods used for risk estimation, we see a similar picture to risk identification. The majority of organizations relies on more informal methods like brainstorming and structured approaches such as \ac{fmea}. Only enterprises operating in strictly regulated domains (e.g. healthcare, utility) apply more formal risk estimation approaches like \ac{fta}, \ac{cca} and \ac{pha}. Tables \ref{tab:results-rt-methods1} and \ref{tab:results-rt-methods2} show the techniques used to prioritize estimated risks and to decide whether risks require treatment at all. Interestingly, dedicated management decisions and cost-benefit analysis are more commonly used than the application of risk acceptance criteria putting greater effort on individual decision processes.

\begin{table}
\caption{Methods applied for risk prioritization ($N=16$)}
\begin{center}
\begin{tabularx}{\linewidth}{Xc}
\toprule
\textbf{Prioritization} & \textbf{Applied by} \\
\toprule
Risk Matrix & $38\%$\\
Risk Priority Number & $38\%$\\
Relative Risk Ranking & $31\%$\\
\midrule
\textit{not disclosed} & $13\%$\\
\bottomrule
\end{tabularx}
\end{center}
\label{tab:results-rt-methods1}
\end{table}%

\begin{table}
\caption{Methods applied for risk treatment decision ($N=16$)}
\begin{center}
\begin{tabularx}{\linewidth}{Xc}
\toprule
\textbf{Decision} & \textbf{Applied by} \\
\toprule
Cost-benefit analysis & $38\%$ \\
Management Decision & $38\%$\\
Risk acceptance criteria & $25\%$\\
\midrule
\textit{not disclosed} & $19\%$\\
\bottomrule
\end{tabularx}
\end{center}
\label{tab:results-rt-methods2}
\end{table}%

\subsubsection*{Security Requirements and Controls}

\begin{figure*}
\begin{center}
\includegraphics[width=\textwidth]{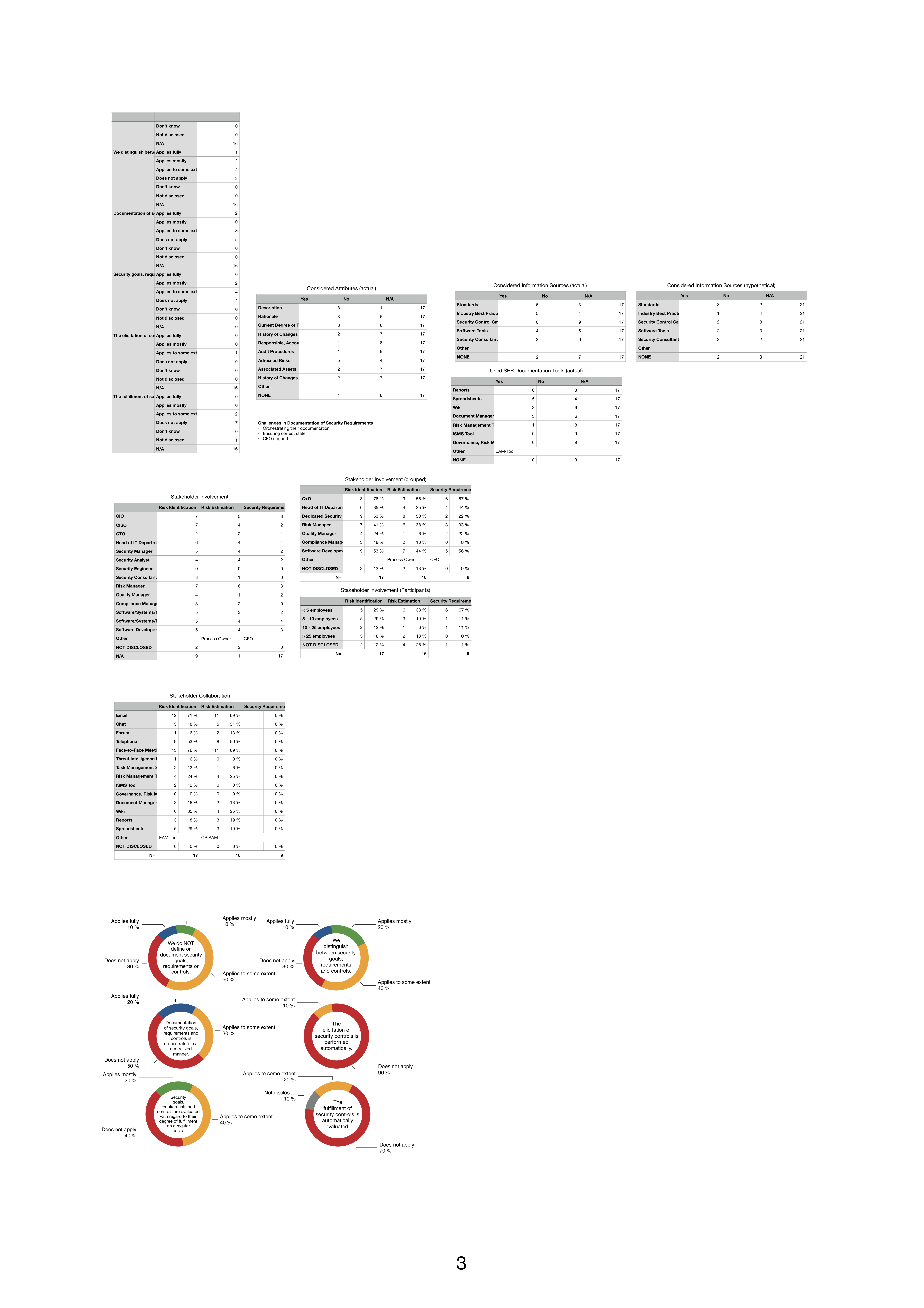}
\caption{State of information security requirements and control practices ($N=10$)}
\label{fig:results-se-practices}
\end{center}
\end{figure*}

The final question block concerning the documentation of security requirements and controls was optional and 10 of 26 responses opted for disclosing their current practices. Figure~\ref{fig:results-se-practices} shows the corresponding results. The majority of enterprises define and document security goals, requirements or controls, with $70\%$ following a hierarchical documentation approach distinguishing at least to some extent between goals, requirements and controls. Documentation is commonly not centralized and re-evaluation of security goals, requirements and controls is only performed on a regular basis by companies operating an  \ac{isms}. Automation techniques in elicitation as well as monitoring of security controls are currently not applied.

The documented attributes and aspects of information security goals, requirements and controls show a high reliance on textual descriptions (cf. Table~\ref{tab:results-se-attributes}). While addressed risks are commonly documented, most respondents refrain from documenting any other rational for security controls or even relating them with the assets they are supposed to protect. Changes of security controls themselves or their fulfillment are of lesser concern as is the dedicated documentation of audit procedures or stakeholder responsibilities.

\begin{table}
\caption{Attributes and aspects documented for security goals, requirements and controls ($N=10$)}
\begin{center}
\begin{tabularx}{\linewidth}{Xc}
\toprule
\textbf{Attribute} & \textbf{Documented by}\\
\toprule
Description & $90\%$ \\
Addressed Risks & $50\%$ \\
Rationale & $30\%$ \\
Associated Assets & $20\%$ \\
History of Changes of the Degree of Fulfillment & $20\%$\\
History of Changes (excl. Degree of Fulfillment) & $20\%$\\
Audit Procedures & $10\%$ \\
Responsible, Accountable, Consulted, Informed (RACI) Stakeholders or Organizational Units & $10\%$ \\
\midrule
\textit{not disclosed} & $20\%$\\
\bottomrule
\end{tabularx}
\end{center}
\label{tab:results-se-attributes}
\end{table}%

\subsection{Stakeholder Involvement and Collaboration Patterns (RQ2)}\label{sec:results:rq2}

In order to investigate stakeholder involvement we asked participants which roles were involved in \ac{isrm} activities and how many people contributed to each step. Table~\ref{tab:results-stakeholder-participation} shows how often each role has been declared to be involved in activities of risk identification, risk estimation and the documentation of security requirements and controls. Responses show that top-management roles (\ac{cio}, \ac{ciso} and to a lesser extent \acs{cto}) are generally involved in all activities, and that the instantiation of dedicated security and risk management roles is common practice. Furthermore we identified quite strong involvement of software development roles. The second part of Table~\ref{tab:results-stakeholder-participation} shows the amount of stakeholders involved in each activity. Although it might be expected that larger organizations with established \ac{isms} invest more manpower in \ac{isrm} activities, responses show that this is not universally true. The companies stating that more than 25 people are involved in any of these actions are all large enterprises with more than 250 employees, already operating an \ac{isms} and working in heavily regulated domains (finance and healthcare). In general, responses show a tendency towards smaller teams performing \ac{isrm} activities, commonly less than 10 people strong even among larger companies.

\begin{table*}
\caption{Stakeholder participation in \ac{isrm} activities}
\begin{center}
\begin{tabularx}{\textwidth}{lXccc}
\toprule
 &  & \textbf{Risk Identification} & \textbf{Risk Estimation} & \textbf{Security Controls}\\
\toprule
\multirow{10}{*}{\textbf{Involved Roles}} & CxO & $76\%$ & $56\%$ & $67\%$ \\
 & Head of IT Department & $35\%$ & $25\%$ & $4\%$ \\
 & Dedicated Security Role & $53\%$ & $50\%$ & $20\%$ \\
 & Risk Manager & $41\%$ & $38\%$ & $30\%$ \\
 & Quality Manager & $24\%$ & $6\%$ & $20\%$ \\
 & Compliance Manager & $18\%$ & $13\%$ & $0\%$ \\
 & Software Development & $53\%$ & $44\%$ & $50\%$ \\
 & Other & --- & Process Owner (1) & CEO (1) \\
\cmidrule{2-5}
 & \textit{not disclosed} & $12\%$ & $13\%$ & $10\%$\\
\bottomrule
\multirow{6}{*}{\textbf{Involved Staff}} & < 5 employees & $29\%$ & $38\%$ & $6\%$ \\
 & 5 - 10 employees & $29\%$ & $19\%$ & $10\%$\\
 & 10 - 25 employees & $12\%$ & $6\%$ & $10\%$\\
 & > 25 employees & $18\%$ & $13\%$ & $0\%$\\
\cmidrule{2-5}
 & \textit{not disclosed} & $12\%$ & $25\%$ & $20\%$\\
\bottomrule
 &  & \footnotesize$N=17$ & \footnotesize$N=16$ & \footnotesize$N=10$ \\
\end{tabularx}
\end{center}
\label{tab:results-stakeholder-participation}
\end{table*}

The question of how stakeholders collaborate in these activities can be conducted in part from previously discussed applied methods. To get a fuller picture, however, we asked dedicated questions regarding their mode of collaboration and how this is technically realized. Table~\ref{tab:results-stakeholder-collaboration} presents the responses to these questions (e.g., RI009 \shortquote{How is the collaboration between stakeholders for information security risk identification designed and which tools are used to document identified risks in your organization?}). We can directly infer a heavy emphasis on direct stakeholder interaction during risk identification and risk estimation, whereas dedicated risk management or \ac{isms} tools play a rather minor role.

\begin{table*}
\caption{Stakeholder collaboration in \ac{isrm} activities}
\begin{center}
\begin{tabularx}{\textwidth}{lXcc}
\toprule
\multicolumn{2}{l}{\textbf{Collaboration Patterns}} & \textbf{Risk Identification} & \textbf{Risk Estimation}\\
\toprule
\multirow{5}{1.8cm}{Direct} & Email & $71\%$ & $69\%$ \\
 & Chat & $18\%$ & $31\%$ \\
 & Forum & $6\%$ & $13\%$ \\
 & Telephone & $53\%$ & $50\%$ \\
 & Face-to-Face Meetings & $76\%$ & $69\%$ \\
\midrule
\multirow{4}{1.8cm}{Tool-based} & Threat Intelligence Sharing Platform & $6\%$ & $0\%$ \\
 & Task Management System & $12\%$ & $6\%$ \\
 & Risk Management Tool & $24\%$ & $31\%$ \\
 & ISMS Tool & $12\%$ & $0\%$ \\
\midrule
\multirow{4}{1.8cm}{Via Shared Documents} & Document Management System & $18\%$ & $13\%$ \\
 & Wiki & $35\%$ & $25\%$ \\
 & Reports & $24\%$ & $19\%$ \\
 & Spreadsheets & $29\%$ & $19\%$ \\
\midrule
\multicolumn{2}{l}{\textit{not disclosed}} & $12\%$ & $13\%$ \\
\bottomrule
 & & \footnotesize$N=17$ & \footnotesize$N=16$ \\
\end{tabularx}
\end{center}
\label{tab:results-stakeholder-collaboration}
\end{table*}%

\subsection{Additional Data Sources and Tools used in Risk Analysis (RQ3)}\label{sec:results:rq3}

Relevant information for \ac{isrm} activities can be retrieved from different data sources and tools. We investigated which data sources are commonly used and which tools enterprises favor. While the previous section emphasized tools used as means to collaborate, this section emphasizes data sources and tools providing input for individual activities and those being used to store respective results and provide them for subsequent activities. 

Considering that the documentation of relevant assets is the vital very first step to conduct a high-quality risk analysis we investigated the tools that are used by enterprises for that activity. Responses indicate that general-purpose documentation tools such as spreadsheets or schematic diagrams and charts are more commonly used than dedicated information security tools (cf. Table~\ref{tab:results-asset-tools}). Specialized documentation and modeling software like CMDBs or EAM tools are typically used in conjunction with general purpose documentation tools. On average an enterprise uses $2.9$ ($s=1.2$) different tool types to document its assets.

\begin{table}
\caption{Tools used to document assets ($N=16$)}
\begin{center}
\begin{tabularx}{\linewidth}{Xc}
\toprule
\textbf{Asset Documentation Tools} & \textbf{Used by}\\
\toprule
Spreadsheets & $50\%$\\
Configuration Management Database (CMDB) & $44\%$\\
Enterprise Architecture Modeling (EAM), Schematic diagrams/charts & $38\%$ \\
ISMS Tool, ISRM Tool & $25\%$ \\
\midrule
\textit{not disclosed} & $25\%$ \\
\bottomrule
\end{tabularx}
\end{center}
\label{tab:results-asset-tools}
\end{table}%

Tables~\ref{tab:results-information-sources-external} and \ref{tab:results-information-sources-internal} list external and internal information sources used during risk identification. The column \emph{Ranking} relates to how often each information source has been named as being one of the three most important information sources for a company. On average, each company used $5.2$ ($s=2.0$) external and $4.6$ ($s=2.4$) internal data sources during risk identification. Enterprises heavily rely on vendor-specific advisories and vulnerability databases. Newspapers, wikis and exploit database, although regularly used, are of lesser importance in risk identification processes. Security policies, incident management as well as (penetration) test reports are the most relevant internal information sources during risk identification. Checklists and security reviews on the other hand are generally considered to be of lesser importance.

\begin{table}
\caption{External information sources used during risk identification ($N=17$)}
\begin{center}
\begin{tabularx}{\linewidth}{Xcc}
\toprule
\textbf{External Information Source} & \textbf{Used by} & \textbf{Ranking}\\
\toprule
Newspapers & $59\%$ & LOW \\
Wikis & $47\%$ & LOW \\
Blogs & $35\%$ & LOW \\
Mailinglists & $65\%$ & MEDIUM\\
Social Media & $29\%$ & LOW \\
Exploit Database & $59\%$ & LOW \\
Vulnerability Database & $65\%$ & HIGH \\
Vendor-specific Advisories & $71\%$ & HIGH \\
Threat Intelligence Sharing Platforms & $29\%$ & LOW \\
Special Interest Groups & $29\%$ & LOW \\
\midrule
\textit{not disclosed} & $6\%$ & \\
\bottomrule
\end{tabularx}
\end{center}
\label{tab:results-information-sources-external}
\end{table}%

\begin{table}
\caption{Internal information sources used during risk identification ($N=17$)}
\begin{center}
\begin{tabularx}{\linewidth}{Xcc}
\toprule
\textbf{Internal Information Source} & \textbf{Used by} & \textbf{Ranking}\\
\toprule
Security Policy & $41\%$ & HIGH \\
Checklists & $47\%$ & LOW \\
Best Practices & $47\%$ & MEDIUM \\
Issue Tracker & $59\%$ & MEDIUM\\
Incident Management & $53\%$ & HIGH \\
Internal (Security) Reviews & $41\%$ & LOW \\
Audit Protocols & $35\%$ & MEDIUM \\
(Penetration) Test Reports & $53\%$ & HIGH \\
Security Monitoring Tools & $53\%$ & MEDIUM \\
\midrule
\textit{not disclosed} & $0\%$ & \\
\bottomrule
\end{tabularx}
\end{center}
\label{tab:results-information-sources-internal}
\end{table}%

The most viable information source for information security requirements and controls are standards and industry best practices (both used by two-thirds of responding companies). Dedicated software tools and the involvement of external security consultants are other applied methods to identify security controls as means of reducing information security risks. The most common tool to document them are written reports and spreadsheets, both a rather static medium. Dedicated document management systems or wikis are of less concern. Only one organization responded that it uses a dedicated risk management tool to document security requirements and controls.

\subsection{Notable Differences between Companies with and without \ac{isms}}\label{sec:results:differences}

Due to the small sample we refrained from quantitative statistic analysis ($N_{\text{ISMS established}} = 10$, $N_{\text{ISMS planned}} = 6$) of differences between those groups. Instead we conducted a qualitative analysis and interpretation of notable differences in responses. Furthermore, we present results from respondents whose organizations did not plan to establish an ISMS or which did not perform certain activities (e.g., enterprises planning to implement an ISMS but not conducting asset documentation yet). Those respondents were guided through the alternate survey path where their perception and general knowledge of \ac{isrm} activities was captured.

In general we observed that \ac{isrm} practices are also performed by organizations not committed to the implementation of an \ac{isms}. Of the seven responses that did not plan to establish an \ac{isms}, three did in fact already document assets, perform risk analysis and document security requirements and controls. Those, however, showed a significant lower maturity in \ac{isrm} practices with less stakeholder involvement and overall smaller scope and documentation discipline.

Organizations who did not document their assets had a slightly different view of what constitutes as a relevant asset for \ac{isrm} activities. Whereas organizational units, suppliers and stakeholders were considered relevant assets by a third of organizations performing this step (cf. Table~\ref{tab:results-artifacts}), they were generally not considered by organizations not performing this practice. Another distinguishing fact was the usage of asset documentation tools. Organizations not documenting assets would favor the use of dedicated documentation tools (e.g., EAM Tool, CMDBs) and not use spreadsheets or schematic diagrams at all. Reality, however, was that actual asset documentation practices primarily relied on the use of these tool types (cf. Section~\ref{sec:results:rq1}).

Respondents from companies which do not perform certain \ac{isrm} practices or are still in the process of planning an \ac{isms} introduction  showed a generally good knowledge of available methods, tools and corresponding standards. Questions targeting preferred \ac{isrm} practices showed a somewhat idealistic view that deviates from actual practice, preferring perceived one-stop solutions (e.g., \ac{isrm} or dedicated modeling tools) over those that are predominantly employed in practice.

\subsection{Perceived Challenges}\label{sec:results:challenges}

Although not the focal point of our investigation, we did ask participants concerning current challenges in information security risk assessment. As initially expected responses to these open questions were sparse but since all of them were given from respondents who were actively involved in their organization's \ac{isms} operation we deem them to be still of value for our overarching research objectives. Reoccurring themes in mentioned challenges were (1) \emph{insufficient management support and availability of stakeholders}, (2) \emph{effort required for formal establishment of \ac{isrm} processes}, and (3) \emph{ensuring that required information is up-to-date}. Apart from (1), these challenges clearly relate to our survey results regarding applied, rather informal \ac{isrm} practices and the neglectable usage of automation facilities or dedicated tools. The lack of top-level management support was mentioned by two respondents from smaller companies (less than 25 employees) operating an \ac{isms}.

\section{Discussion}\label{sec:conclusions}

The conducted study investigated \ac{isrm} practices applied in larger organization-wide information security management settings with an emphasis on organizations operating or planning to introduce an \ac{isms}. We constructed a conceptual model to explore the status quo of risk management activities, involved stakeholders, their collaboration patterns as well as utilized tools and data sources. Our analysis used data provided by 26 participants who shared information pertaining aspects of our conceptual model. In this section, we will discuss the results, their potential implications with regard to our research objectives and potential limitations of our research.

\subsection{Interpretation of Results}\label{sec:discussion}

In general our respondents closely followed respective standards and best practices when choosing methods and providing documentation artifacts within their \ac{isrm} activities. Starting with the first step, the documentation of assets, results show that relevant aspects and groups of assets are well represented in general. However, a greater emphasis on purely technical aspects, especially within small or medium sized enterprises is present. 
This further indicates that more complex aspects in information security risk management such as securing the availability of critical organizational knowledge are currently underrepresented. 
The preferred use of general purpose documentation tools such as spreadsheets or diagrams in favor of tools dedicated to manage asset inventories has multiple implications. It leads to higher manual involvement and introduces potential errors due to non-timely updates, even when asset inventories are mostly managed in a centralized fashion. Furthermore, the granularity and quality of asset catalogs can be improved by a more thorough consideration of individual assets instead of groups of assets and the documentation of potential dependencies between assets. 

A similar reliance on manual processes was found throughout all subsequent risk assessment activities. The amount of involved stakeholders and the heterogeneity of involved business roles would ideally require the use of focused practices and close monitoring which was not the case. Instead neither applied risk assessment methods nor means of stakeholder collaboration can be considered satisfactory in terms of repeatability and overall output quality. Our results point to the typical risk identification practice involving face-to-face meetings supplemented by telephone calls and emails between stakeholders who rely on brainstorming techniques and checklists to identify information security risks. Accordingly, these practices limit the traceability and documentation of decisions which might result in a lack of transparency. That is not only far from utilizing readily available more structured or even formal approaches, but also burdens involved stakeholders with scheduling issues for required meetings or results in non-availability of key stakeholders (cf. Section~\ref{sec:results:challenges}). The primarily qualitative risk estimation approaches applied by respondents bear the same pitfalls especially when paired with predominantly unstructured documentation of the results of performed \ac{isrm} activities. Overall, our results suggest that \ac{isrm} practices are conducted in a fashion that strongly impedes their reliability -- especially when key personell is replaced or otherwise not available. 

Considering the heterogeneity of involved stakeholders it is not surprising that a diverse set of information sources are used during \ac{isrm} activities as well. Information sources perceived as more important typically provide data points that are directly applicable to \ac{isrm} activities such as vendor-specific advisories which can easily be distilled to retrieve the assets being subject to a certain vulnerability and additionally provide a preemptive risk analysis. Still, less structured information sources (e.g., newspaper articles, mailing lists, wikis) are commonly used by stakeholders. Reliably analyzing these resources for relevant information is an elaborate task that would not only involve extensive scanning of unstructured material but in addition require stakeholders to convey retrieved information to their organizational setting as well. Moreover, research on shadow threat intelligence showed that the informal use of less structured information might result in several risks like limited traceability, information loss or waste of resources \citep{Sauerwein:2018sh}. This situation might be improved through the introduction of NLP-supported threat intelligence (sharing) platforms that provide relevant information tailored to the specific information security needs of an organization and the respective demands of involved stakeholders.

The definition, documentation and subsequent management of security requirements and controls leaves room for improvement as well. Especially technical security controls could favor more automation and shorter re-assessment cycles regarding the review of their fulfillment. Various solutions are available to automatically monitor the fulfillment of certain security controls and integrating them in organizational \ac{isrm} practices would greatly benefit the timeliness of available compliance information. Especially in light of the thorough documentation practices applied for security controls -- they are commonly documented with links to addressed risks -- this could generally improve other \ac{isrm} activities as well and could potentially lead to a more timely re-evaluation of information security risks.

\subsection{FAIR Guidelines}

Concerning potential future \ac{isrm} research and framework development in organizational \ac{isrm} and \acp{isms} we propose the following guidelines for better applicability of developed solutions in light of the reported status quo:
\begin{itemize}
\item\hspace{-1.25em}\circledb{\hspace{0.1em}F}\textbf{avor structured over strictly formal approaches.} \ac{isrm} practices are currently dominated by informal approaches. A direct leap to more demanding formal practices -- especially for risk analysis in general organizational settings -- will not be widely adopted. Research should thus focus on providing structured approaches with easy-to-follow guidelines and clear instructions for result documentation.
\item\hspace{-1.25em}\circledb{A}\textbf{ddress heterogeneous stakeholder landscape.} Information security initiatives heavily rely on inclusion and collaboration of different stakeholders (security experts, process owners, etc.) from various domains. Any successful approach will have to proactively address collaboration patterns and potential issues due to differences in stakeholder knowledge and expertise. 
\item\hspace{-1.25em}\circledb{\hspace{0.2em}I\hspace{0.2em}}\textbf{ncorporate established documentation practices.} Independent of applied practices and standards, \ac{isrm} activities utilize various information sources and storage facilities, in many cases relying on general purpose documentation tools. Enterprises will not be forced to abandon these established documentation practices and new approaches should aim at seamless integration of what is present and working.
\item\hspace{-1.25em}\circledb{R}\textbf{espect scarceness of resources.} Operating an \ac{isms} is a costly business endeavor due to high reliance on manual decision processes and overall lack of automation. The complex enterprise spanning scope thus requires provision of technical, financial and human resources often outside an enterprises' dedicated security organization. Risk management approaches should thus provide a transparent cost-benefit model to show that scarce resources are used beneficially and that automation facilities are effectively implemented.
\end{itemize}
These guidelines were derived from the previously presented interpretation of results and cover the major reoccurring themes found in our survey responses.
We argue that by following our proposed guidelines, researchers can develop more directly applicable \ac{isrm} solutions that substantially improve the current state of practice without overburdening enterprises with additional efforts. While the process discipline and overall maturity of \ac{ism} is expected to raise in the coming years, solutions adhering our guidelines should aid organizational transitions until formal approaches are becoming the norm as security experts have been advocating for decades.

\subsection{Implications}\label{sec:implications}

Our results provide researchers with a more comprehensive picture of currently applied \ac{isrm} practices in organizations. Since our investigation emphasized organization-wide practices applied within \acp{isms} we explored what techniques enterprises use in heterogeneous intra-organizational settings instead of secluded risk management activities performed in specialized application domains. We provide new insight into involved roles and collaboration patterns within these activities including preferred tool usage and incorporated information sources.

We provide the basis for the development of \ac{isrm} practices that are less disruptive of the current organizational practice and thus have a greater chance to be actually adopted by enterprises. Furthermore, our proposed guidelines can help researchers to successfully transfer available conceptual tools or approaches to larger, practical, closer-to-life settings for evaluation purposes and critical reflection of their own works applicability.

Three common deficiencies in \ac{isrm} practices were identified by \cite{Webb:2014hg}: (1) information security risk identification is commonly perfunctory, (2) information security risks are commonly estimated with little reference to the organization's actual situation, and (3) information Security risk assessment is commonly performed on an intermitted, non-historical basis. The general focus on technical aspects in asset documentation practices as well as the insufficient consideration of dependencies between assets and risks shown by our results largely support deficiency 1. A similar issue regarding proper identification and inventory of information technology assets was stated in challenge 1 by~\cite{Fenz:2014cur}. 
The organizational reality regarding risk estimation as illustrated by our study responses highlights the general refusal of applying quantitative approaches in favor of rather interpretative practices with high reliance on stakeholders' expertise and limited reproducibility of results. This compares to deficiency 2 by~\cite{Webb:2014hg} and is partially captured by challenge 3 regarding failed predictions in risk in~\cite{Fenz:2014cur}. 
In alignment with deficiency 3 our responses support the notion of primarily intermitted non-historical risk assessments. This is obvious in the malpractice of performing risk management activities irregularly with two or more years in between cycles. Which is made even worse by manual and non-formal risk assessment methods generally used in practice and a heavy reliance on general purpose documentation tools with questionable abilities to reliably portrait historical developments to involved stakeholders. 

Regarding \ac{isrm} research in general, \cite{Wangen:2013wb} constitute various challenges, repeatedly stating a common lack of empirical research and good data, especially concerning the validation and verification of existing methods. We argue that without a better alignment between proposed risk management approaches and the current industrial practice -- as well as organizational capabilities -- this gap can not be sufficiently addressed on a broader scale. Our survey design and results contribute a solid basis for further empirical research and provides a viable starting point to tailor academic efforts to a wider base of organizations. 

Overall, with our contribution, we support researchers with a solid picture of the current industrial practice and a repeatable survey instrument to periodically re-assess the status quo. Additionally, the proposed guidelines should help academic endeavors and information security practitioners alike to develop applicable solutions for the iterative improvement of current \ac{isrm} practices.

\subsection{Limitations}\label{sec:limitations}

Our survey has been developed in accordance with respected best practice guidelines and we performed several measures to control the validity of presented results as outlined in Section~\ref{sec:method}. However, there are still limitations present that need further consideration.

Most notably, the number of responses that we received and that could ultimately be used is limited, thus potentially affecting the external validity of our findings. It is, however, reasonably high for the performed qualitative analysis and descriptive statistics presented in this work. A larger set of responses would ultimately yield more reliable results and allow the application of quantitative statistic analysis methods. 

Currently, we do not have any means to establish the representativeness of our study population since no reliable figures regarding \ac{isrm} or \ac{isms} adoption in the DACH region are available. This, naturally, restricts the generalizability of our findings. We presume for example that the actual business domain and corresponding regulatory demands influences applied \ac{isrm} practices and their actual implementation. Since only a fraction of responses belongs to these domains we cannot derive any definitive conclusions. Medium sized (50 - 250 employees) and large companies (> 250 employees) make up the majority of responses. This prohibits us from generalizing our conclusions to smaller enterprises.

The qualitative data analysis applied to analyze responses might pose a threat to the reliability of derived results. While the data set was small enough to be processed in that fashion, the identification of patterns in responses is arguably a creative process and thus potentially influenced by expectations and prior experiences of involved researchers. We mitigated this threat by involving multiple researchers in data analysis.  

The design of our survey instrument raises potential threats to construct validity of our results. First, the online survey was aimed at a maximum duration of 20 minutes and was thus comparably more extensive than other information security surveys. This poses the risk of participants preemptively aborting our survey or quickly ending it via extensive use of default options and ignoring non-mandatory questions. Responses showed that participants commonly required around 19 minutes to complete the survey and that those who prematurely aborted the questionnaire did so during early questions regarding demographics. We additionally analyzed the timespan participants required for individual questions as well as the usage of default answer options. From that we could not identify any tendency that later question were not as thoroughly considered and answered as early ones. However, based on our results we plan to streamline the survey instrument for future iterations by merging several questions regarding stakeholder collaboration and tool usage. 

The second potential issue regarding the survey instrument arises from ambiguities in applied terms and definitions. We counteracted this largely by choosing the well-established ISO 27k family of standards as referential basis and by providing additional clarifications and definitions with each question. Furthermore, participants were given ample opportunity to express deviations via commendatory input blocks present in most questions. Together with the results of a conducted pilot survey and subsequent interviews with pilot participants we are confident that the design of our online questionnaire did not negatively influence the validity of our research.

Finally, our decision to use the ISO 27k family of standards as referential basis of our survey instrument poses a threat to the validity of our findings and has a potentially negative influence the generalizability of our results. This affects primarily participants not familiar with the ISO standards that are operating an \ac{isms} based on conceptually different approaches such as the BSI IT Baseline Protection Methodology~\citep{bsi:2017ms} or the NIST \ac{rmf}~\citep{NIST:sp-800-37-r2} for example. These standards put a stronger emphasis on documentation and classification of processed information and utilized systems together with the provision of baseline security controls to reach the desired level of information security. Typical \ac{isrm} activities are thus not procedurally represented in the same fashion as in the ISO 27k family of standards which might confound the aforementioned group of study participants.
The received responses and comments from participants -- even those that aborted the study and were subsequently discarded for analysis -- did not provide any indication that this has been an issue.

\section{Summary and Future Work}\label{sec:summary}

We presented an exploratory survey concerning the status quo of risk management practices in information security. We based our investigation on a conceptual model exploring different aspects such as the applied \ac{isrm} methodology, patterns of stakeholder collaboration, utilized tools as well as involved information sources and considered documentation artifacts. We identified that the current state of practice in the DACH region has a strong emphasis on manual data collection, direct stakeholder communication and non-formal approaches for risk identification and estimation as well as complex but unstructured decision processes. In addition, our findings suggest that the use of general purpose documentation tools is preferred over using dedicated risk management or \ac{isms} tools. Finally, we derived guidelines for the development of \ac{isrm} frameworks better suiting the current state of practice and thereby enhancing their chance for contemporary industrial application. 

Our own research in the area of continuous information security management and the development of the corresponding ADAMANT framework already benefited from early findings \citep{Brunner:2018wx,Brunner:2019fb} and we will continue to improve our approach. Following our guidelines we will emphasize the integration of established documentation practices and investigate potential means to reduce efforts for the introduction and operation of \acp{isms} using ADAMANT. In addition, we are currently designing a study to empirically analyze the proposed guidelines and their potential impact on stakeholder perception of \ac{isrm} approaches regarding their usefulness, applicability and implementation effort. Future research will additionally target small and medium-sized enterprises, investigating different \ac{isrm} approaches to establish and support smaller scale \acp{isms}.

Another important direction of future research will be the iterative enhancement and replication of the presented study. We thus want to invite information security researchers to join us in our effort to draw a more precise picture of the current state of \ac{isrm} practices used by enterprises to manage information security -- not only in the DACH region but also on a global scale.

\section*{Acknowledgments}
This work has been partially sponsored and supported by the Austrian Ministry for Transport, Innovation and Technology by Projects \shortquote{SALSA} (Project-No. 855383), \shortquote{Q-WEST} (Project-No. 858574), \shortquote{Digital Tourism Experts} (Project-No. 866123) and \shortquote{Digitalisierung und Sicherheit} (Project-No. 872612).


\bibliographystyle{cas-model2-names}

\balance

\bibliography{references}

\vskip3pt

\bio{}
\textbf{Michael Brunner} is a doctoral student at the Department of Computer Science at the University of Innsbruck, Austria. His primary research interests are information security management systems, risk assessment techniques and empirical research of tool-based information security management approaches. Furthermore, he investigates the unification of security and safety aspects of cyber-physical systems and the application of continuous information security management practices in this domain. He transfers his results into industrial settings as consultant and lecturer in research transfer projects.
\endbio

\bio{}
\textbf{Clemens Sauerwein} is a post-doctoral researcher at the Department of Computer Science at the University of Innsbruck, Austria. His research interests include information security risk management, cyber threat intelligence sharing, empirical studies in the field of information security risk management and information systems. He works in close collaboration with industry and transfers his results into practice as a consultant and a member of a security interest group.
\endbio

\bio{}
\textbf{Michael Felderer} is a professor in software engineering at the Department of Computer Science at the University of Innsbruck, Austria and a guest professor at the Blekinge Institute of Technology, Sweden. He holds a Ph.D. and a habilitation degree in computer science. His research interests in software and security engineering include software and security testing, empirical methods in software and security engineering, software and security processes, software analytics, risk management, requirements engineering, model engineering, and data-driven engineering. He works in close collaboration with industry and is a regular speaker at industrial conferences.
\endbio

\bio{}
\textbf{Ruth Breu} is head of the Department of Computer Science at the University of Innsbruck and head of the research group Quality Engineering. She is expert in the areas of Requirements Engineering, Security Engineering and Enterprise Architecture Management. Together with her team she develops tool-based methods for in- formation security management and IT asset documentation with a high degree of automation, collaboration support and situation- awareness. Ruth is co-author of more than 150 international publications and contributor to the scientific community as editor, conference and PC chair.
\endbio

\onecolumn

\normalsize
\appendix

\section*{Appendix}

The following presentation of the survey instrument is content-complete and a faithful recreation of the online questionnaire used für our research. It does not aim at accurately portraying the look-and-feel of the resulting webpages which is hardly possible in print.


\section{Survey Instrument: RiskFlows Exploratory Study (Online)}\label{app:survey}

The RiskFlows explorative study investigates the current practice and shortcomings in information security risk management workflows. The study is conducted by the $<$AUTHOR-AFFILIATION$>$. The findings will provide ample information on viable approaches for novel risk-driven information security management workflows that will additionally address the areas of risk treatment and monitoring.

Thank you for considering to participate in our study. If you work in the line of strategic or operational information security (risk) management you are the prime candidate for this questionnaire. We will ask specifics about the current state of affairs regarding information security management and risk assessment in your company.

All responses are stored anonymously, none of the responses will be connected to identifying information, the results will be used for statistical purposes and will be reported only in aggregated form. The survey will take approximately 20 minutes to complete.

\subsection{General}

This group contains general questions regarding your enterprise and your organizational role.

\subsubsection*{DE001: What is your organizational role?} (Multiple\textsuperscript{+})
\vspace{-1em}\begin{multicols}{2}
\noindent$\square$~Chief Information Officer\\
$\square$~Chief Information Security Officer\\
$\square$~Chief Technology Officer\\
$\square$~Head of IT Department\\
$\square$~Security Manager\\
$\square$~Security Analyst\\
$\square$~Security Engineer\\
$\square$~Security Consultant\\
$\square$~Risk Manager\\
$\square$~Quality Manager\\
$\square$~Compliance Manager\\
$\square$~Software/Systems/Network Engineer\\
$\square$~Software/Systems/Network Architect\\
$\square$~Software Developer\\ \\
Other: \_\_\_\_\_\_\_\_\_\_\_\_\_\_\_\_\_\_\_\_\_\_\_\_
\end{multicols}

\subsubsection*{DE002: Which of the following personal certifications and qualifications do you have?} (Multiple\textsuperscript{+})
\vspace{-1em}\begin{multicols}{2}
\noindent$\square$~CISM\\
$\square$~CISSP\\
$\square$~CEH\\
$\square$~CCNP\\
$\square$~CISA\\
$\square$~University degree in Computer Science\\
$\square$~University degree in Information Systems\\
$\square$~University degree in Business or Economics\\ \\
Other: \_\_\_\_\_\_\_\_\_\_\_\_\_\_\_\_\_\_\_\_\_\_\_\_
\end{multicols}

\subsubsection*{DE003: How many years of professional expierience in the area of information security do you have?} (Numeric)

\subsubsection*{DE004: What type of industry is your organization in?} (Single\textsuperscript{+})
\vspace{-1em}\begin{multicols}{2}
\noindent$\circ$~Information Technology\\
$\circ$~Services\\
$\circ$~Manufacturing\\
$\circ$~Retail\\
$\circ$~Finance and Insurance\\
$\circ$~Public Administration\\ \\
Other: \_\_\_\_\_\_\_\_\_\_\_\_\_\_\_\_\_\_\_\_\_\_\_\_
\end{multicols}

\subsubsection*{DE005: What is the size of your organization?} (Single)
\vspace{-1em}\begin{multicols}{2}
\noindent$\circ$~$<$ 10 employees\\
$\circ$~10 - 50 employees\\
$\circ$~50 - 250 employees\\
$\circ$~$>$ 250 employees
\end{multicols}

\pagebreak

\subsubsection*{DE006: Where is your organization located?} (Single\textsuperscript{*})
\vspace{-1em}\begin{multicols}{3}
\noindent$\circ$~Germany\\
$\circ$~Austria\\
$\circ$~Switzerland\\
$\circ$~Liechtenstein\\
$\circ$~Other EU Country\\
$\circ$~Non EU Country
\end{multicols}
\vspace{-1em}
\noindent Subsidiaries in: \_\_\_\_\_\_\_\_\_\_\_\_\_\_\_\_\_\_\_\_\_\_\_\_\_\_\_\_\_\_\_\_\_\_\_\_\_\_\_\_\_\_\_\_\_\_\_\_

\subsubsection*{DE007: Which ICT Standards and frameworks are (partially) implemented in your orgnization?} (Multiple\textsuperscript{+})
\vspace{-1em}\begin{multicols}{2}
\noindent$\square$~ISO/IEC 27000 family of Standards\\
$\square$~COBIT\\
$\square$~ITIL\\
$\square$~ValIT\\
$\square$~BSI Baseline Protection Methodology\\ \\ \\
Other: \_\_\_\_\_\_\_\_\_\_\_\_\_\_\_\_\_\_\_\_\_\_\_\_
\end{multicols}

\subsection{Information Security Management}

This short group of questions aims at understanding basic notions of information security management in your organization. According to the international ISO 27000 standard an Information Security Management System (ISMS) "[...] consists of the policies, procedures, guidelines, and associated resources and activities, collectively managed by an organization, in the pursuit of protecting its information assets."

\subsubsection*{IM001: Do you operate an \acl{isms}?} (Single\textsuperscript{+})
\vspace{-1em}
\begin{multicols}{2}
\noindent$\circ$~No, and we do not plan to implement an ISMS\\
\noindent$\circ$~No, but we plan to implement an ISMS\\ \\
\noindent$\circ$~Yes, we operate a non-certified ISMS\\
\noindent$\circ$~Yes, we operate a certified ISMS (please provide certification information in the comment)
\end{multicols}
\vspace{-1em}
\noindent Comment: \_\_\_\_\_\_\_\_\_\_\_\_\_\_\_\_\_\_\_\_\_\_\_\_\_\_\_\_\_\_\_\_\_\_\_\_\_\_\_\_\_\_\_\_\_\_\_\_

\subsubsection*{IM002: What is the most important driver for Information Security Management in your organization?} (Single\textsuperscript{+})
\vspace{-1em}
\begin{multicols}{2}
\noindent$\circ$~Higher degree of information security\\
\noindent$\circ$~Customer demands\\
\noindent$\circ$~Shareholder demands\\
\noindent$\circ$~Regulatory demands\\ \\
Other: \_\_\_\_\_\_\_\_\_\_\_\_\_\_\_\_\_\_\_\_\_\_\_\_
\end{multicols}

\subsubsection*{IM003: Which Information Security Risk Management (ISRM) methodology does your organization apply?} (Multiple\textsuperscript{+})
\vspace{-1em}\begin{multicols}{2}
\noindent$\square$~ISO 27005\\
\noindent$\square$~ISO 31000\\
\noindent$\square$~BSI IT Baseline Protection Methodology\\
\noindent$\square$~NIST SP 800 30\\
\noindent$\square$~CRAMM\\
\noindent$\square$~OCTAVE\\
\noindent$\square$~AS/NZS 4360\\
\noindent$\square$~CRISAM\\ \\
Other: \_\_\_\_\_\_\_\_\_\_\_\_\_\_\_\_\_\_\_\_\_\_\_\_
\end{multicols}

\subsubsection*{IM004: How often does your organization conduct an information security risk management cycle?} (Single)
\vspace{-1em}\begin{multicols}{1}
\noindent$\circ$~Never\\
\noindent$\circ$~At least twice a year\\
\noindent$\circ$~At least once a year\\
\noindent$\circ$~At least once every two years\\
\noindent$\circ$~Occasionally (more than two years between cycles)\\ \\
\noindent$\circ$~No answer
\end{multicols}

\subsubsection*{IM005: Which events additionally trigger information security risk management activities in your organization?} (Multiple\textsuperscript{+})
\vspace{-1em}\begin{multicols}{1}
\noindent$\square$~Change of the operational Environment (Process, IT Infrastructure, ...)\\
\noindent$\square$~Reported Security Incident or vulnerability (external)\\
\noindent$\square$~Attack of IT infrastructure, applications or services (internal)\\
\noindent$\square$~Internal Audit\\
\noindent$\square$~External Audit\\ \\ \\
Other: \_\_\_\_\_\_\_\_\_\_\_\_\_\_\_\_\_\_\_\_\_\_\_\_
\end{multicols}

\pagebreak

\subsubsection*{IM002\textsubscript{alt}: What do you consider the most important driver for Information Security Management?} (Single\textsuperscript{+})
\vspace{-1em}\begin{multicols}{2}
\noindent$\circ$~Higher degree of information security\\
\noindent$\circ$~Customer demands\\
\noindent$\circ$~Shareholder demands\\
\noindent$\circ$~Regulatory demands\\ \\
Other: \_\_\_\_\_\_\_\_\_\_\_\_\_\_\_\_\_\_\_\_\_\_\_\_
\end{multicols}

\subsubsection*{IM003\textsubscript{alt}: Which Information Security Risk Management (ISRM) methodology do you know?} (Multiple\textsuperscript{+})
\vspace{-1em}\begin{multicols}{2}
\noindent$\square$~ISO 27005\\
\noindent$\square$~ISO 31000\\
\noindent$\square$~BSI IT Baseline Protection Methodology\\
\noindent$\square$~NIST SP 800 30\\
\noindent$\square$~CRAMM\\
\noindent$\square$~OCTAVE\\
\noindent$\square$~AS/NZS 4360\\
\noindent$\square$~CRISAM\\ \\
Other: \_\_\_\_\_\_\_\_\_\_\_\_\_\_\_\_\_\_\_\_\_\_\_\_
\end{multicols}

\subsection{Considered Assets}

This group of questions asks specifics about the way your organization documents and manages assets relevant for information security. An Asset is "[...] any tangible or intangible thing or characteristic that has value to an organization". ISO 27001 demands that all assets "[...] associated with information and information processing facilities shall be identified and an inventory of these assets shall be drawn up and maintained."

Even if your organization does not operate an ISMS or does not perform Information Security Risk Management (ISRM), you might keep records of relevant assets (such as IT services, IT infrastructure elements, data, business processes).

\subsubsection*{AS000: Do you document assets in your organization in any way, shape or form?} (Yes/No)

\subsubsection*{AS001: Rate the following statements with regard to the asset documentation \\of your organization.} (Rating)
\vspace{-4em}
\begin{center}
\begin{tabularx}{\textwidth}{Xcccccc}
 & \rotatebox{90}{Applies fully} & \rotatebox{90}{Applies mostly} & \rotatebox{90}{Applies to some extent} & \rotatebox{90}{Does not apply} & \rotatebox{90}{Do not know} & \rotatebox{90}{No answer} \\
\toprule
Asset documentation is orchestrated in a centralized manner.  & $\circ$ & $\circ$ & $\circ$ & $\circ$ & $\circ$ & $\circ$ \\
\midrule
We keep records of all individual assets. & $\circ$ & $\circ$ & $\circ$ & $\circ$ & $\circ$ & $\circ$ \\
\midrule
Asset documentation is always up-to-date. & $\circ$ & $\circ$ & $\circ$ & $\circ$ & $\circ$ & $\circ$ \\
\midrule
We document dependencies between assets. & $\circ$ & $\circ$ & $\circ$ & $\circ$ & $\circ$ & $\circ$ \\
\midrule
Asset discovery/documentation is performed automatically. & $\circ$ & $\circ$ & $\circ$ & $\circ$ & $\circ$ & $\circ$ \\
\bottomrule
\end{tabularx}
\end{center}

\subsubsection*{AS002: Which assets are considered for the ISMS and ISRM activities in your organization?} (Multiple\textsuperscript{+})
\vspace{-1em}\begin{multicols}{2}
\noindent$\square$~Business Processes\\
\noindent$\square$~Organizational Units\\
\noindent$\square$~Suppliers\\
\noindent$\square$~Stakeholders\\
\noindent$\square$~IT Services\\
\noindent$\square$~Applications\\
\noindent$\square$~Cloud Services (external)\\
\noindent$\square$~Server Hardware\\
\noindent$\square$~Workstations\\
\noindent$\square$~POI/POS Terminals\\
\noindent$\square$~Network Infrastructure\\
\noindent$\square$~Premises\\
\noindent$\square$~Rooms\\ \\ \\
Other: \_\_\_\_\_\_\_\_\_\_\_\_\_\_\_\_\_\_\_\_\_\_\_\_
\end{multicols}

\pagebreak

\subsubsection*{AS003: Which tools do you use manage the considered assets in your organization?} (Multiple\textsuperscript{+})
\vspace{-1em}\begin{multicols}{2}
\noindent$\square$~Enterprise Architecture Management (EAM)\\
\noindent$\square$~Configuration Management Database (CMDB)\\
\noindent$\square$~ISMS Tool\\
\noindent$\square$~ISRM Tool\\
\noindent$\square$~Spreadsheets\\
\noindent$\square$~Schematic diagrams/charts\\ \\
Other: \_\_\_\_\_\_\_\_\_\_\_\_\_\_\_\_\_\_\_\_\_\_\_\_
\end{multicols}

\subsubsection*{AS002\textsubscript{alt}: Which assets would you consider relevant for ISMS and ISRM activities?} (Multiple\textsuperscript{+})
\vspace{-1em}\begin{multicols}{2}
\noindent$\square$~Business Processes\\
\noindent$\square$~Organizational Units\\
\noindent$\square$~Suppliers\\
\noindent$\square$~Stakeholders\\
\noindent$\square$~IT Services\\
\noindent$\square$~Applications\\
\noindent$\square$~Cloud Services (external)\\
\noindent$\square$~Server Hardware\\
\noindent$\square$~Workstations\\
\noindent$\square$~POI/POS Terminals\\
\noindent$\square$~Network Infrastructure\\
\noindent$\square$~Premises\\
\noindent$\square$~Rooms\\ \\ \\
Other: \_\_\_\_\_\_\_\_\_\_\_\_\_\_\_\_\_\_\_\_\_\_\_\_
\end{multicols}

\subsubsection*{AS003\textsubscript{alt}: How would you prefer to manage considered assets?} (Multiple\textsuperscript{+})
\vspace{-1em}\begin{multicols}{2}
\noindent$\square$~Enterprise Architecture Management (EAM)\\
\noindent$\square$~Configuration Management Database (CMDB)\\
\noindent$\square$~ISMS Tool\\
\noindent$\square$~ISRM Tool\\
\noindent$\square$~Spreadsheets\\
\noindent$\square$~Schematic diagrams/charts\\ \\
Other: \_\_\_\_\_\_\_\_\_\_\_\_\_\_\_\_\_\_\_\_\_\_\_\_
\end{multicols}

\subsection{Risk Identification}

This group of questions asks specifics about the Information Security Risk Management (ISRM) approach at your organization. In particular these questions target the the way how risks are identified at your organization as part of your ISMS or ISRM initiative. According to ISO 27005 Risk Identification is "[...] the process to find, list and characterize elements of risk".

\subsubsection*{RI000: Does your organization perform Information Security Risk Management (ISRM) or related activities?} (Yes/No)\\
\vspace{-2em}\begin{center}$\circ$~Yes\hspace{10em}$\circ$~No\end{center}
\small{Please choose \textbf{Yes}, if your organization has established ISRM processes or conducts any kinds of tasks in support of information security risk management (identification, estimation or evaluation of information security risks, treatment of information security risks, assessment of protection levels for assets, etc.).\\
Only choose \textbf{No}, if your organization does NOT perform any kind of Information Security Risk Management (ISRM).}\normalsize

\subsubsection*{RI001: Rate the following statements with regard to the information security\\ risk identification approach of your organization.} (Rating)
\vspace{-4em}
\begin{center}
\begin{tabularx}{\textwidth}{Xcccccc}
 & \rotatebox{90}{Applies fully} & \rotatebox{90}{Applies mostly} & \rotatebox{90}{Applies to some extent} & \rotatebox{90}{Does not apply} & \rotatebox{90}{Do not know} & \rotatebox{90}{No answer} \\
\toprule
Risk identification is performed automatically.  & $\circ$ & $\circ$ & $\circ$ & $\circ$ & $\circ$ & $\circ$ \\
\midrule
Every relevant security risk is identified in a timely fashion. & $\circ$ & $\circ$ & $\circ$ & $\circ$ & $\circ$ & $\circ$ \\
\midrule
Sharing of relevant security information is conducted via a formal process. & $\circ$ & $\circ$ & $\circ$ & $\circ$ & $\circ$ & $\circ$ \\
\midrule
Relevant security information is automatically preprocessed and filtered for conducting risk identification. & $\circ$ & $\circ$ & $\circ$ & $\circ$ & $\circ$ & $\circ$ \\
\midrule
Exchange of security information with other organizations and individuals has been beneficial for risk identification. & $\circ$ & $\circ$ & $\circ$ & $\circ$ & $\circ$ & $\circ$ \\
\bottomrule
\end{tabularx}
\end{center}

\subsubsection*{RI002: Which aspects are considered for information security risk identification in your organization?} (Multiple\textsuperscript{+})
\vspace{-1em}\begin{multicols}{2}
\noindent$\square$~Assets\\
\noindent$\square$~Threats\\
\noindent$\square$~Vulnerabilities\\
\noindent$\square$~Existing or Planned Controls and Countermeasures\\
\noindent$\square$~Security Incidents\\
\noindent$\square$~Consequences\\
\noindent$\square$~Security Goals\\
\noindent$\square$~Required Protection Level of Assets\\ \\
Other: \_\_\_\_\_\_\_\_\_\_\_\_\_\_\_\_\_\_\_\_\_\_\_\_
\end{multicols}

\subsubsection*{RI003: Which EXTERNAL information sources are used for information security risk identification in your organization?} (Multiple\textsuperscript{+})
\vspace{-1em}\begin{multicols}{2}
\noindent$\square$~Newspapers\\
\noindent$\square$~Wikis\\
\noindent$\square$~Blogs\\
\noindent$\square$~Mailinglists\\
\noindent$\square$~Social Media\\
\noindent$\square$~Exploit Database\\
\noindent$\square$~Vulnerability Database\\
\noindent$\square$~Vendor-specific Advisories\\
\noindent$\square$~Threat Intelligence Sharing Platforms\\
\noindent$\square$~Special Interest Groups\\ \\
Other: \_\_\_\_\_\_\_\_\_\_\_\_\_\_\_\_\_\_\_\_\_\_\_\_
\end{multicols}

\subsubsection*{RI004: Which are the three most important EXTERNAL information sources for information security risk identification in your organization?} (Ranking)

\subsubsection*{RI005: Which INTERNAL information sources are used for information security risk identification in your organization?} (Multiple\textsuperscript{+})
\vspace{-1em}\begin{multicols}{2}
\noindent$\square$~Security Policy\\
\noindent$\square$~Checklists\\
\noindent$\square$~Best Practices\\
\noindent$\square$~Issue Tracker\\
\noindent$\square$~Incident Management\\
\noindent$\square$~Internal (Security) Reviews\\
\noindent$\square$~Audit Protocols\\
\noindent$\square$~(Penetration) Test Reports\\
\noindent$\square$~Security Monitoring Tools\\ \\
Other: \_\_\_\_\_\_\_\_\_\_\_\_\_\_\_\_\_\_\_\_\_\_\_\_
\end{multicols}

\subsubsection*{RI006: Which are the three most important INTERNAL information sources for information security risk identification in your organization?} (Ranking)

\subsubsection*{RI007: Which stakeholders are involved in information security risk identification in your organization?} (Multiple\textsuperscript{+})
\vspace{-1em}\begin{multicols}{2}
\noindent$\square$~Chief Information Officer\\
$\square$~Chief Information Security Officer\\
$\square$~Chief Technology Officer\\
$\square$~Head of IT Department\\
$\square$~Security Manager\\
$\square$~Security Analyst\\
$\square$~Security Engineer\\
$\square$~Security Consultant\\
$\square$~Risk Manager\\
$\square$~Quality Manager\\
$\square$~Compliance Manager\\
$\square$~Software/Systems/Network Engineer\\
$\square$~Software/Systems/Network Architect\\
$\square$~Software Developer\\ \\
Other: \_\_\_\_\_\_\_\_\_\_\_\_\_\_\_\_\_\_\_\_\_\_\_\_
\end{multicols}

\subsubsection*{RI008: How many stakeholders are involved in information security risk identification within your organization?} (Single\textsuperscript{+})
\vspace{-1em}
\begin{multicols}{2}
\noindent$\circ$~$<$ 5 Employees\\
\noindent$\circ$~5 - 10 employees\\
\noindent$\circ$~20 - 25 employees\\
\noindent$\circ$~$>$ 25 employees\\ \\
\noindent$\circ$~No answer
\end{multicols}
\vspace{-1em}
\noindent Comment: \_\_\_\_\_\_\_\_\_\_\_\_\_\_\_\_\_\_\_\_\_\_\_\_\_\_\_\_\_\_\_\_\_\_\_\_\_\_\_\_\_\_\_\_\_\_\_\_

\pagebreak

\subsubsection*{RI009: How is the collaboration between stakeholders for information security risk identification designed and which tools are used to document identified risks in your organization?} (Multiple\textsuperscript{+})
\vspace{-1em}\begin{multicols}{2}
\noindent$\square$~Email\\
\noindent$\square$~Chat\\
\noindent$\square$~Forum\\
\noindent$\square$~Telephone
\noindent$\square$~Face-to-Face Meetings\\
\noindent$\square$~Threat Intelligence Sharing Platform\\
\noindent$\square$~Task Management System\\
\noindent$\square$~Risk Management Tool\\
\noindent$\square$~ISMS Tool\\
\noindent$\square$~Governance, Risk Management and Compliance (GRC) Tool\\
\noindent$\square$~Document Management System\\
\noindent$\square$~Wiki\\
\noindent$\square$~Reports\\
\noindent$\square$~Spreadsheets\\ \\
Other: \_\_\_\_\_\_\_\_\_\_\_\_\_\_\_\_\_\_\_\_\_\_\_\_
\end{multicols}

\subsubsection*{RI010: Which methods are used for identifying information security risks in your organization?} (Multiple\textsuperscript{+})
\vspace{-1em}\begin{multicols}{2}
\noindent$\square$~Brainstorming\\
\noindent$\square$~Checklists\\
\noindent$\square$~Structured What If Technique (SWIFT)\\
\noindent$\square$~Preliminary Hazard Analysis (PHA)\\
\noindent$\square$~Failure Mode and Effect Analysis (FMEA)\\
\noindent$\square$~Hazard and Operability Study (HAZOP)\\
\noindent$\square$~Cause and Consequence Analysis (CCA)\\
\noindent$\square$~Reliability Availability, Maintainability Analysis (RAM)\\
\noindent$\square$~CCTA Risk Analysis and Management Method (CRAMM)\\
\noindent$\square$~Information Security Risk Analysis Method (ISRAM)\\
\noindent$\square$~CORAS Method\\
\noindent$\square$~Consultative, Objective and Bi-functional Risk Analysis (COBRA)\\
\noindent$\square$~Operationally Critical Threat, Asset, and Vulnerability Evaluation (OCTAVE)\\ \\
Other: \_\_\_\_\_\_\_\_\_\_\_\_\_\_\_\_\_\_\_\_\_\_\_\_
\end{multicols}

\subsubsection*{RI011: What are the most pressing challenges during information security risk identification for your organization?} (Open)\\
\noindent\_\_\_\_\_\_\_\_\_\_\_\_\_\_\_\_\_\_\_\_\_\_\_\_\_\_\_\_\_\_\_\_\_\_\_\_\_\_\_\_\_\_\_\_\_\_\_\_\_\_\_\_\_\_\_\_\_\_\_\_\_\_\_\_\_\_\_\_\_\_\_\_\_\_\_\_\_\_\_\_\_\_\\
\_\_\_\_\_\_\_\_\_\_\_\_\_\_\_\_\_\_\_\_\_\_\_\_\_\_\_\_\_\_\_\_\_\_\_\_\_\_\_\_\_\_\_\_\_\_\_\_\_\_\_\_\_\_\_\_\_\_\_\_\_\_\_\_\_\_\_\_\_\_\_\_\_\_\_\_\_\_\_\_\_\_\\
\_\_\_\_\_\_\_\_\_\_\_\_\_\_\_\_\_\_\_\_\_\_\_\_\_\_\_\_\_\_\_\_\_\_\_\_\_\_\_\_\_\_\_\_\_\_\_\_\_\_\_\_\_\_\_\_\_\_\_\_\_\_\_\_\_\_\_\_\_\_\_\_\_\_\_\_\_\_\_\_\_\_\\

\subsubsection*{RI003\textsubscript{alt}: Which EXTERNAL information sources supporting the identification of information security risks do you know?} (Multiple\textsuperscript{+})
\vspace{-1em}\begin{multicols}{2}
\noindent$\square$~Newspapers\\
\noindent$\square$~Wikis\\
\noindent$\square$~Blogs\\
\noindent$\square$~Mailinglists\\
\noindent$\square$~Social Media\\
\noindent$\square$~Exploit Database\\
\noindent$\square$~Vulnerability Database\\
\noindent$\square$~Vendor-specific Advisories\\
\noindent$\square$~Threat Intelligence Sharing Platforms\\
\noindent$\square$~Special Interest Groups\\ \\
Other: \_\_\_\_\_\_\_\_\_\_\_\_\_\_\_\_\_\_\_\_\_\_\_\_
\end{multicols}

\subsubsection*{RI005\textsubscript{alt}: Which INTERNAL information sources supporting the identification of information security risks do you know?} (Multiple\textsuperscript{+})
\vspace{-1em}\begin{multicols}{2}
\noindent$\square$~Security Policy\\
\noindent$\square$~Checklists\\
\noindent$\square$~Best Practices\\
\noindent$\square$~Issue Tracker\\
\noindent$\square$~Incident Management\\
\noindent$\square$~Internal (Security) Reviews\\
\noindent$\square$~Audit Protocols\\
\noindent$\square$~(Penetration) Test Reports\\
\noindent$\square$~Security Monitoring Tools\\ \\ \\
Other: \_\_\_\_\_\_\_\_\_\_\_\_\_\_\_\_\_\_\_\_\_\_\_\_
\end{multicols}

\pagebreak

\subsubsection*{RI009\textsubscript{alt}: How would you prefer to design the collaboration between stakeholders for information security risk identification and which tools would you prefer to use to document identified risks?} (Multiple\textsuperscript{+})
\vspace{-1em}\begin{multicols}{2}
\noindent$\square$~Email\\
\noindent$\square$~Chat\\
\noindent$\square$~Forum\\
\noindent$\square$~Telephone
\noindent$\square$~Face-to-Face Meetings\\
\noindent$\square$~Threat Intelligence Sharing Platform\\
\noindent$\square$~Task Management System\\
\noindent$\square$~Risk Management Tool\\
\noindent$\square$~ISMS Tool\\
\noindent$\square$~Governance, Risk Management and Compliance (GRC) Tool\\
\noindent$\square$~Document Management System\\
\noindent$\square$~Wiki\\
\noindent$\square$~Reports\\
\noindent$\square$~Spreadsheets\\ \\
Other: \_\_\_\_\_\_\_\_\_\_\_\_\_\_\_\_\_\_\_\_\_\_\_\_
\end{multicols}

\subsubsection*{RI010\textsubscript{alt}: Which methods for identifying information security risks do you know?} (Multiple\textsuperscript{+})
\vspace{-1em}\begin{multicols}{2}
\noindent$\square$~Brainstorming\\
\noindent$\square$~Checklists\\
\noindent$\square$~Structured What If Technique (SWIFT)\\
\noindent$\square$~Preliminary Hazard Analysis (PHA)\\
\noindent$\square$~Failure Mode and Effect Analysis (FMEA)\\
\noindent$\square$~Hazard and Operability Study (HAZOP)\\
\noindent$\square$~Cause and Consequence Analysis (CCA)\\
\noindent$\square$~Reliability Availability, Maintainability Analysis (RAM)\\
\noindent$\square$~CCTA Risk Analysis and Management Method (CRAMM)\\
\noindent$\square$~Information Security Risk Analysis Method (ISRAM)\\
\noindent$\square$~CORAS Method\\
\noindent$\square$~Consultative, Objective and Bi-functional Risk Analysis (COBRA)\\
\noindent$\square$~Operationally Critical Threat, Asset, and Vulnerability Evaluation (OCTAVE)\\ \\
Other: \_\_\_\_\_\_\_\_\_\_\_\_\_\_\_\_\_\_\_\_\_\_\_\_
\end{multicols}

\subsection{Risk Estimation, Evaluation and Treatment}

This group of questions asks specifics about the Information Security Risk Management (ISRM) approach at your organization. In particular these questions target the the way how risks are estimated/evaluated at your organization as part of your ISMS or ISRM initiative and how your organization decides which treatment options of risks are pursued. According to ISO 27005 Risk Estimation is "[...] the process to assign values to the probability and consequence of a risk" whereas Risk Evaluation is defined as  "[...] the process of comparing the results of risk analysis [...] to determine whether the risk and/or its magnitude is acceptable or tolerable."

\subsubsection*{RE001: Which stakeholders are involved in information security risk estimation and evaluation in your organization?} (Multiple\textsuperscript{+})
\vspace{-1em}\begin{multicols}{2}
\noindent$\square$~Chief Information Officer\\
$\square$~Chief Information Security Officer\\
$\square$~Chief Technology Officer\\
$\square$~Head of IT Department\\
$\square$~Security Manager\\
$\square$~Security Analyst\\
$\square$~Security Engineer\\
$\square$~Security Consultant\\
$\square$~Risk Manager\\
$\square$~Quality Manager\\
$\square$~Compliance Manager\\
$\square$~Software/Systems/Network Engineer\\
$\square$~Software/Systems/Network Architect\\
$\square$~Software Developer\\ \\
Other: \_\_\_\_\_\_\_\_\_\_\_\_\_\_\_\_\_\_\_\_\_\_\_\_
\end{multicols}

\subsubsection*{RE002: How many stakeholders are involved in the estimation of information security risks in your organization?} (Single\textsuperscript{+})
\vspace{-1em}
\begin{multicols}{2}
\noindent$\circ$~$<$ 5 Employees\\
\noindent$\circ$~5 - 10 employees\\
\noindent$\circ$~20 - 25 employees\\
\noindent$\circ$~$>$ 25 employees\\ \\
\noindent$\circ$~No answer
\end{multicols}
\vspace{-1em}
\noindent Comment: \_\_\_\_\_\_\_\_\_\_\_\_\_\_\_\_\_\_\_\_\_\_\_\_\_\_\_\_\_\_\_\_\_\_\_\_\_\_\_\_\_\_\_\_\_\_\_\_

\pagebreak

\subsubsection*{RE003: How is the collaboration between stakeholders for information security risk estimation and evaluation designed and which tools are used to document the risk estimation results in your organization?} (Multiple\textsuperscript{+})
\vspace{-1em}\begin{multicols}{2}
\noindent$\square$~Email\\
\noindent$\square$~Chat\\
\noindent$\square$~Forum\\
\noindent$\square$~Telephone
\noindent$\square$~Face-to-Face Meetings\\
\noindent$\square$~Threat Intelligence Sharing Platform\\
\noindent$\square$~Task Management System\\
\noindent$\square$~Risk Management Tool\\
\noindent$\square$~ISMS Tool\\
\noindent$\square$~Governance, Risk Management and Compliance (GRC) Tool\\
\noindent$\square$~Document Management System\\
\noindent$\square$~Wiki\\
\noindent$\square$~Reports\\
\noindent$\square$~Spreadsheets\\ \\
Other: \_\_\_\_\_\_\_\_\_\_\_\_\_\_\_\_\_\_\_\_\_\_\_\_
\end{multicols}

\subsubsection*{RE004: Your organization's risk estimation approach is ...} (Single\textsuperscript{+})
\vspace{-1em}
\begin{multicols}{2}
\noindent$\circ$~Qualitative (Subjective and scale-based, e.g., critical, high, medium, low)\\
\noindent$\circ$~ Semi-Quantitative\\
\noindent$\circ$~Quantitative (Calculated, e.g., expected annual financial loss)\\
\noindent$\circ$~Don't know\\ \\
\noindent$\circ$~No answer
\end{multicols}
\vspace{-1em}
\noindent Comment: \_\_\_\_\_\_\_\_\_\_\_\_\_\_\_\_\_\_\_\_\_\_\_\_\_\_\_\_\_\_\_\_\_\_\_\_\_\_\_\_\_\_\_\_\_\_\_\_

\subsubsection*{RE005: Which methods are used for estimating information security risks in your organization?} (Multiple\textsuperscript{+})
\vspace{-1em}\begin{multicols}{2}
\noindent$\square$~Brainstorming\\
\noindent$\square$~Structured What If Technique (SWIFT)\\
\noindent$\square$~Preliminary Hazard Analysis (PHA)\\
\noindent$\square$~Failure Mode and Effect Analysis (FMEA)\\
\noindent$\square$~Hazard and Operability Study (HAZOP)\\
\noindent$\square$~Fault Tree Analysis (FTA)\\
\noindent$\square$~Event Tree Analysis (ETA)\\
\noindent$\square$~Cause and Consequence Analysis (CCA)\\
\noindent$\square$~Monte-Carlo Simulation\\
\noindent$\square$~Reliability Availability, Maintainability Analysis (RAM)\\
\noindent$\square$~CCTA Risk Analysis and Management Method (CRAMM)\\
\noindent$\square$~Information Security Risk Analysis Method (ISRAM)\\
\noindent$\square$~CORAS Method\\
\noindent$\square$~Consultative, Objective and Bi-functional Risk Analysis (COBRA)\\
\noindent$\square$~Operationally Critical Threat, Asset, and Vulnerability Evaluation (OCTAVE)\\ \\
Other: \_\_\_\_\_\_\_\_\_\_\_\_\_\_\_\_\_\_\_\_\_\_\_\_
\end{multicols}

\subsubsection*{RE006: Rate the following statements with regard to the information security\\  risk estimation approach in your organization.} (Rating)
\vspace{-4em}
\begin{center}
\begin{tabularx}{\textwidth}{Xcccccc}
 & \rotatebox{90}{Applies fully} & \rotatebox{90}{Applies mostly} & \rotatebox{90}{Applies to some extent} & \rotatebox{90}{Does not apply} & \rotatebox{90}{Do not know} & \rotatebox{90}{No answer} \\
\toprule
Dependencies between ASSETS (e.g., between business processes and the IT infrastructure to deliver them) are considered when estimating security risks. & $\circ$ & $\circ$ & $\circ$ & $\circ$ & $\circ$ & $\circ$ \\
\midrule
Dependencies between RISKS (e.g., between risk of a reduced availability of a virtualized server and the risk of reduced availability of the hardware node it is running on) are considered when estimating security risks. & $\circ$ & $\circ$ & $\circ$ & $\circ$ & $\circ$ & $\circ$ \\
\midrule
The estimation of the PROBABILITY of a single risk to materialize takes related risks into account. & $\circ$ & $\circ$ & $\circ$ & $\circ$ & $\circ$ & $\circ$ \\
\midrule
The estimation of the LIKELIHOOD of a single risk to materialize takes related risks into account.& $\circ$ & $\circ$ & $\circ$ & $\circ$ & $\circ$ & $\circ$ \\
\midrule
Risk estimation is performed automatically. & $\circ$ & $\circ$ & $\circ$ & $\circ$ & $\circ$ & $\circ$ \\
\bottomrule
\end{tabularx}
\end{center}

\pagebreak

\subsubsection*{RE007: Which methods are used to prioritize estimated information security risks in your organization?} (Multiple\textsuperscript{+})
\vspace{-1em}\begin{multicols}{2}
\noindent$\square$~Risk Matrix\\
\noindent$\square$~Risk Priority Number\\
\noindent$\square$~Relative Risk Ranking\\
\noindent$\square$~Failure Mode Effect and Criticality Analysis (FMECA)\\
Other: \_\_\_\_\_\_\_\_\_\_\_\_\_\_\_\_\_\_\_\_\_\_\_\_
\end{multicols}

\subsubsection*{RE008: What are the most pressing challenges during information security risk estimation and evaluation for your organization?} (Open)\\
\noindent\_\_\_\_\_\_\_\_\_\_\_\_\_\_\_\_\_\_\_\_\_\_\_\_\_\_\_\_\_\_\_\_\_\_\_\_\_\_\_\_\_\_\_\_\_\_\_\_\_\_\_\_\_\_\_\_\_\_\_\_\_\_\_\_\_\_\_\_\_\_\_\_\_\_\_\_\_\_\_\_\_\_\\
\_\_\_\_\_\_\_\_\_\_\_\_\_\_\_\_\_\_\_\_\_\_\_\_\_\_\_\_\_\_\_\_\_\_\_\_\_\_\_\_\_\_\_\_\_\_\_\_\_\_\_\_\_\_\_\_\_\_\_\_\_\_\_\_\_\_\_\_\_\_\_\_\_\_\_\_\_\_\_\_\_\_\\
\_\_\_\_\_\_\_\_\_\_\_\_\_\_\_\_\_\_\_\_\_\_\_\_\_\_\_\_\_\_\_\_\_\_\_\_\_\_\_\_\_\_\_\_\_\_\_\_\_\_\_\_\_\_\_\_\_\_\_\_\_\_\_\_\_\_\_\_\_\_\_\_\_\_\_\_\_\_\_\_\_\_\\

\subsubsection*{RE009: How do you decide whether information security risks are accepted or treated in your organization?} (Multiple\textsuperscript{+})
\vspace{-1em}\begin{multicols}{2}
\noindent$\square$~Defined risk acceptance criteria\\
\noindent$\square$~Cost-benefit analysis of treatment options\\
\noindent$\square$~Management Decision\\
Other: \_\_\_\_\_\_\_\_\_\_\_\_\_\_\_\_\_\_\_\_\_\_\_\_
\end{multicols}

\subsubsection*{RE003\textsubscript{alt}: How would you prefer to design the collaboration between stakeholders for information security risk estimation and evaluation and which tools would you prefer to use to document risk estimation results?} (Multiple\textsuperscript{+})
\vspace{-1em}\begin{multicols}{2}
\noindent$\square$~Email\\
\noindent$\square$~Chat\\
\noindent$\square$~Forum\\
\noindent$\square$~Telephone
\noindent$\square$~Face-to-Face Meetings\\
\noindent$\square$~Threat Intelligence Sharing Platform\\
\noindent$\square$~Task Management System\\
\noindent$\square$~Risk Management Tool\\
\noindent$\square$~ISMS Tool\\
\noindent$\square$~Governance, Risk Management and Compliance (GRC) Tool\\
\noindent$\square$~Document Management System\\
\noindent$\square$~Wiki\\
\noindent$\square$~Reports\\
\noindent$\square$~Spreadsheets\\ \\
Other: \_\_\_\_\_\_\_\_\_\_\_\_\_\_\_\_\_\_\_\_\_\_\_\_
\end{multicols}

\subsubsection*{RE004\textsubscript{alt}:  I would prefer to utilize a ... risk estimation approach.} (Single\textsuperscript{+})
\vspace{-1em}
\begin{multicols}{2}
\noindent$\circ$~Qualitative (Subjective and scale-based, e.g., critical, high, medium, low)\\
\noindent$\circ$~ Semi-Quantitative\\
\noindent$\circ$~Quantitative (Calculated, e.g., expected annual financial loss)\\
\noindent$\circ$~Don't know\\ \\
\noindent$\circ$~No answer
\end{multicols}
\vspace{-1em}
\noindent Comment: \_\_\_\_\_\_\_\_\_\_\_\_\_\_\_\_\_\_\_\_\_\_\_\_\_\_\_\_\_\_\_\_\_\_\_\_\_\_\_\_\_\_\_\_\_\_\_\_

\subsubsection*{RE005\textsubscript{alt}:  Which methods for estimating information security risks do you know?} (Multiple\textsuperscript{+})
\vspace{-1em}\begin{multicols}{2}
\noindent$\square$~Brainstorming\\
\noindent$\square$~Structured What If Technique (SWIFT)\\
\noindent$\square$~Preliminary Hazard Analysis (PHA)\\
\noindent$\square$~Failure Mode and Effect Analysis (FMEA)\\
\noindent$\square$~Hazard and Operability Study (HAZOP)\\
\noindent$\square$~Fault Tree Analysis (FTA)\\
\noindent$\square$~Event Tree Analysis (ETA)\\
\noindent$\square$~Cause and Consequence Analysis (CCA)\\
\noindent$\square$~Monte-Carlo Simulation\\
\noindent$\square$~Reliability Availability, Maintainability Analysis (RAM)\\
\noindent$\square$~CCTA Risk Analysis and Management Method (CRAMM)\\
\noindent$\square$~Information Security Risk Analysis Method (ISRAM)\\
\noindent$\square$~CORAS Method\\
\noindent$\square$~Consultative, Objective and Bi-functional Risk Analysis (COBRA)\\
\noindent$\square$~Operationally Critical Threat, Asset, and Vulnerability Evaluation (OCTAVE)\\ \\ \\
Other: \_\_\_\_\_\_\_\_\_\_\_\_\_\_\_\_\_\_\_\_\_\_\_\_
\end{multicols}

\subsection{Security Goals, Requirements and Controls}
This additional group of questions asks specifics about the way that security goals, requirements and controls are defined and documented at your organization. According to ISO 27000 a security control is defined as "[...] measure that is modifying risk", a requirement is a "[...] need or expectation that is stated, generally implied or obligatory" and goals typically describe "[...] results to be achieved".

\subsubsection*{SE000: May we ask you additional questions regarding the definition and documentation of security goals, requirements and controls.} (Yes/No)\\
\vspace{-1em}\begin{center}$\circ$~Yes\hspace{10em}$\circ$~No\end{center}
\small{Thank you for answering the previous questions. If you can spare another 5 minutes, we would like to ask you about the definition and documentation of security goals, requirements and controls.}\normalsize

\subsubsection*{SE001: Rate the following statements with regard to the elicitation and \\documentation of security goals, requirements and controls within your \\organization.} (Rating)
\vspace{-5em}
\begin{center}
\begin{tabularx}{\textwidth}{Xcccccc}
 & \rotatebox{90}{Applies fully} & \rotatebox{90}{Applies mostly} & \rotatebox{90}{Applies to some extent} & \rotatebox{90}{Does not apply} & \rotatebox{90}{Do not know} & \rotatebox{90}{No answer} \\
\toprule
We do NOT define or document security goals, requirements or controls.& $\circ$ & $\circ$ & $\circ$ & $\circ$ & $\circ$ & $\circ$ \\
\midrule
We distinguish between security goals, requirements and controls.& $\circ$ & $\circ$ & $\circ$ & $\circ$ & $\circ$ & $\circ$ \\
\midrule
Documentation of security goals, requirements and controls is orchestrated in a centralized manner. & $\circ$ & $\circ$ & $\circ$ & $\circ$ & $\circ$ & $\circ$ \\
\midrule
Security goals, requirements and controls are evaluated with regard to their degree of fulfillment on a regular basis. & $\circ$ & $\circ$ & $\circ$ & $\circ$ & $\circ$ & $\circ$ \\
\midrule
The elicitation of security controls is performed automatically. & $\circ$ & $\circ$ & $\circ$ & $\circ$ & $\circ$ & $\circ$ \\\midrule
The fulfillment of security controls is automatically evaluated. & $\circ$ & $\circ$ & $\circ$ & $\circ$ & $\circ$ & $\circ$ \\
\bottomrule
\end{tabularx}
\end{center}

\subsubsection*{SE002: Which stakeholders are involved in defining security requirements and controls to treat relevant security risks in your organization?} (Multiple\textsuperscript{+})
\vspace{-1em}\begin{multicols}{2}
\noindent$\square$~Chief Information Officer\\
$\square$~Chief Information Security Officer\\
$\square$~Chief Technology Officer\\
$\square$~Head of IT Department\\
$\square$~Security Manager\\
$\square$~Security Analyst\\
$\square$~Security Engineer\\
$\square$~Security Consultant\\
$\square$~Risk Manager\\
$\square$~Quality Manager\\
$\square$~Compliance Manager\\
$\square$~Software/Systems/Network Engineer\\
$\square$~Software/Systems/Network Architect\\
$\square$~Software Developer\\ \\
Other: \_\_\_\_\_\_\_\_\_\_\_\_\_\_\_\_\_\_\_\_\_\_\_\_
\end{multicols}

\subsubsection*{SE003: How many stakeholders are involved in the definition of security requirements or controls in your organization?} (Single\textsuperscript{+})
\vspace{-1em}
\begin{multicols}{2}
\noindent$\circ$~$<$ 5 Employees\\
\noindent$\circ$~5 - 10 employees\\
\noindent$\circ$~20 - 25 employees\\
\noindent$\circ$~$>$ 25 employees\\ \\
\noindent$\circ$~No answer
\end{multicols}
\vspace{-1em}
\noindent Comment: \_\_\_\_\_\_\_\_\_\_\_\_\_\_\_\_\_\_\_\_\_\_\_\_\_\_\_\_\_\_\_\_\_\_\_\_\_\_\_\_\_\_\_\_\_\_\_\_

\subsubsection*{SE004: Which EXTERNAL information sources are used in finding appropriate security goals, requirements and controls in your organization?} (Multiple\textsuperscript{+})
\vspace{-1em}\begin{multicols}{2}
\noindent$\square$~Standards\\
\noindent$\square$~Industry Best Practices\\
\noindent$\square$~Security Control Catalogues\\
\noindent$\square$~Software Tools\\
\noindent$\square$~Security Consultants\\ \\ \\
Other: \_\_\_\_\_\_\_\_\_\_\_\_\_\_\_\_\_\_\_\_\_\_\_\_
\end{multicols}

\pagebreak

\subsubsection*{SE005: Which tools are used to document security requirements and controls in your organization?} (Multiple\textsuperscript{+})
\vspace{-1em}\begin{multicols}{2}
\noindent$\square$~Reports\\
\noindent$\square$~Spreadsheets\\
\noindent$\square$~Wiki\\
\noindent$\square$~Document Management System\\
\noindent$\square$~Risk Management Tool\\
\noindent$\square$~ISMS Tool\\
\noindent$\square$~Governance, Risk Management and Compliance (GRC) Tool\\ \\
Other: \_\_\_\_\_\_\_\_\_\_\_\_\_\_\_\_\_\_\_\_\_\_\_\_
\end{multicols}

\subsubsection*{SE006: Which attributes and aspects are documented for security goals, requirements and controls in your organization?} (Multiple\textsuperscript{+})
\vspace{-1em}\begin{multicols}{2}
\noindent$\square$~Description\\
\noindent$\square$~Rationale\\
\noindent$\square$~Current Degree of Fulfillment\\
\noindent$\square$~History of Changes of the Degree of Fulfillment\\
\noindent$\square$~Responsible, Accountable, Consulted, Informed (RACI) Stakeholders or Organizational Units\\
\noindent$\square$~Audit Procedures\\
\noindent$\square$~Adressed Risks\\
\noindent$\square$~Associated Assets\\
\noindent$\square$~History of Changes (excl. Degree of Fulfillment)\\ \\ \\
Other: \_\_\_\_\_\_\_\_\_\_\_\_\_\_\_\_\_\_\_\_\_\_\_\_
\end{multicols}

\subsubsection*{SE007: What are the most pressing challenges regarding the definition and documentation of security requirements and controls for your organization?} (Open)\\
\noindent\_\_\_\_\_\_\_\_\_\_\_\_\_\_\_\_\_\_\_\_\_\_\_\_\_\_\_\_\_\_\_\_\_\_\_\_\_\_\_\_\_\_\_\_\_\_\_\_\_\_\_\_\_\_\_\_\_\_\_\_\_\_\_\_\_\_\_\_\_\_\_\_\_\_\_\_\_\_\_\_\_\_\\
\_\_\_\_\_\_\_\_\_\_\_\_\_\_\_\_\_\_\_\_\_\_\_\_\_\_\_\_\_\_\_\_\_\_\_\_\_\_\_\_\_\_\_\_\_\_\_\_\_\_\_\_\_\_\_\_\_\_\_\_\_\_\_\_\_\_\_\_\_\_\_\_\_\_\_\_\_\_\_\_\_\_\\
\_\_\_\_\_\_\_\_\_\_\_\_\_\_\_\_\_\_\_\_\_\_\_\_\_\_\_\_\_\_\_\_\_\_\_\_\_\_\_\_\_\_\_\_\_\_\_\_\_\_\_\_\_\_\_\_\_\_\_\_\_\_\_\_\_\_\_\_\_\_\_\_\_\_\_\_\_\_\_\_\_\_\\


\section{Invitation Mail}\label{app:invitation}

Dear $<$PARTICIPANT-NAME$>$,
\\\\
We are writing to request your participation in our explorative risk management study of information security risk and compliance experts in the D.A.CH. area. This study investigates the current practice and shortcomings in information security risk management workflows. The study is conducted by the $<$AUTHOR-AFFILIATION$>$ and the findings will provide ample information on viable approaches for novel risk-driven information security management workflows that will additionally address the areas of risk treatment and monitoring.
\\\\
All responses are stored anonymously, none of the responses will be connected to identifying information, the results will be used for statistical purposes and will be reported only in aggregated form. The survey will take approximately 20 minutes to complete.
\\\\
To participate, please click on the following link:
$<$SURVEY-URL$>$
\\\\
If you have any questions about this survey, or difficulty in accessing the site or completing the survey, please contact $<$AUTHOR-URL$>$.
\\\\
Thank you in advance for participating in this survey.
\\\\
Sincerely,
$<$AUTHOR$>$


\end{document}